\documentclass[10pt]{article}
\usepackage{amssymb}
\usepackage{latexsym}

\def\be{\begin{equation}}
\def\ee{\end{equation}}
\def\bea{\begin{eqnarray}}
\def\eea{\end{eqnarray}}
\def\Tr{{\rm \,Tr\,}}
\def\tr{{\rm \,tr\,}}
\def\Tr{{\rm \,Tr\,}}
\def\tr{{\rm \,tr\,}}
\def\d{{\,\rm d}}

\def\r{{\bf r}}
\def\p{{\bf p}}

\def\x{{\bf x}}
\def\y{{\bf y}}

\def\veps{\varepsilon}
\def\h2m{\frac{\hbar^2}{2m}}
\def\p0{{P_{\beta H^0_N}}}
\def\pna{{P_{N,a}}}
\def\pnan{{P_{N,a_N}}}
\def\pnanl{{P_{N,a_N}\left(\frac{n_0}{N}\geq\lambda_N\right)}}
\def\pnlam{{P_{N,a_N}\left(\frac{n_0}{N}\geq\lambda\right)}}

\newtheorem{theorem}{Theorem}

\newtheorem{lemma}{Lemma}[section]

\newtheorem{prop}{Proposition}[section]
\newtheorem{coro}{Corollary}[section]

\begin{document}

\title{\large\bf Normal and generalized Bose condensation in traps: One dimensional
examples}
\author{Andr\'as S\"ut\H o\\
Research Institute for Solid State Physics and Optics\\ Hungarian Academy of
Sciences \\ P. O. Box 49, H-1525 Budapest\\ Hungary\\
E-mail: suto@szfki.hu}
\date{}
\maketitle
\thispagestyle{empty}
\begin{abstract}
\noindent
We prove the following results. (i)
One-dimensional Bose gases which interact via
unscaled integrable pair interactions
and are
confined in an external potential increasing faster than quadratically undergo a
complete generalized Bose-Einstein condensation (BEC)
at any temperature, in the sense that a macroscopic number of particles are distributed on a $o(N)$
number of one-particle states.
(ii) In a one dimensional harmonic trap the replacement of
the oscillator frequency $\omega$
by $\omega\ln N/N$ gives rise to a phase transition
at $a\equiv \hbar\omega\beta=1$ in the noninteracting
gas. For $a<1$ the limit
distribution of $n_0/N^a$ is exponential and $\langle n_0\rangle/N^a\to 1$.
For $a>1$ there is
BEC with a condensate density $\langle n_0\rangle/N\to 1-a^{-1}$.
For $a\geq 1$, $(\ln N/N)(n_0-\langle n_0\rangle)$
is asymptotically distributed following Gumbel's
law.
For any $a>0$ the free energy is
$-(\pi^2/6\beta a)N/\ln N+o(N/\ln N)$,
with no singularity at $a=1$.
(iii) In Model (ii) both above and below the critical temperature
the gas undergoes a complete generalized BEC, thus providing a
coexistence of ordinary and generalized condensates below the critical point.
(iv) Adding an interaction $\langle U_N\rangle=o(N\ln N)$ to Model (ii) we prove that
a complete generalized BEC occurs for any $\beta>0$.

\vspace{2mm}
\noindent
PACS numbers: 0530J, 0550, 7510J

\vspace{2mm}
\noindent
{\bf KEY WORDS:} trapped Bose gas; one dimension; generalised Bose-Einstein condensation; interactions
\end{abstract}


\section{Introduction}
The idea of Bose-Einstein condensation (BEC) in the ground state of
a one-dimensional interacting Bose gas first appeared in Girardeau's
1960 paper about the $\delta$-gas in the impenetrable limit
\cite{Gir}. Based on an approximate calculation, Girardeau suggested that there is no
BEC in the ground state but, instead, there is a complete
generalized BEC (GBEC) in the sense that
\be\label{1}
\lim_{s\to 0}\lim_{N\to\infty}\frac{1}{N}\sum_{|k|<s\rho}\langle n_k\rangle=1\ .
\ee
Here $N$ is the number of particles, $\langle n_k\rangle$ is the ground state
expectation value of the occupation number of the one-particle state
$\sim\exp(ikx)$ and $\rho$ is the density. Later Schultz \cite{Sch}
disproved this conjecture by showing that the double limit in (\ref{1})
yields zero. A subsequent work by Lenard \cite{Len} on the momentum distribution
confirmed this conclusion. Somewhat later
the absence of BEC \cite{H} (see \cite{BM} for a rigorous proof) and that of GBEC \cite{G2} were
shown in one- and two-dimensional homogenous systems at positive temperatures $T$.
Bogoliubov's inequality \cite{B} which these results are based on becomes trivial at $T=0$,
and BEC or GBEC in the ground state of the {\em soft-}
$\delta$-gas has been a most interesting open problem ever since the Bethe-Ansatz
solution by Lieb and Liniger was published \cite {LL}.
Different approximate methods, as an
effective long-range theory \cite{Hal}, expansion to first order
in $1/c$ where $c$ is the prefactor of $\delta$ \cite{CTW}
and conformal field theoric approach \cite{KBI},
predict an algebraic decay of the
ground state expectation value of the boson field operator $a^*(x)a(0)$,
incompatible with an off-diagonal long-range order.
Also, some arguments \cite{Col,PS} excluding the spontaneous breakdown of a
continuous symmetry in the ground state of certain one-dimensional
systems probably apply to the $\delta$-gas.
On the other hand, no conjecture seems to exist concerning the possibility
of GBEC.

BEC can occur in one dimension if the gas is confined in an external
potential.
In \cite{S} it was shown that there is BEC at any finite temperature if
the bosons are in a fixed (not $N$-dependent) external
potential and interact via a suitably scaled pair interaction,
\be\label{scale}
u_N(\x)=b_N u(\alpha_N \x)\ .
\ee
Here $u$ is an integrable positive function (a soft $\delta$ in one dimension is
allowed)
and $b_N$ and $\alpha_N$ have to be chosen so that
\be\label{uN}
\int u_N\d\x\leq C/N
\ee
for some finite $C$. This result is valid in any dimension $d\geq 1$ and imposes
\be\label{bnalfn}
b_N\leq C\alpha_N^d/N.
\ee
Interactions satisfying this condition cover
in any dimension a whole range
from mean-field types ($\alpha_N$ and, thus, $b_N$ tending to
zero as $N$ goes to infinity) to sharply concentrated ones ($\alpha_N$
and $b_N$ going to infinity).
Even the mean-field
type interactions are nontrivial if the confining potential is
locally bounded, thus allowing particles to decrease their interaction by
increasing their separation, or if $u(\x)$ tends to infinity as $\x$ goes to zero.

For the noninteracting gas
Bose condensation in a fixed trap is a pure ground state phenomenon.
Let $H^0=-\h2m\Delta+V$ where $V$ is chosen in such a way that
$\tr e^{-\beta H^0}<\infty$ for any $\beta>0$. (By Symanzik's version \cite{Sym} of the
Golden-Thompson-Symanzik inequality 
\be
\tr e^{-\beta H^0}\leq \lambda_\beta^{-d}\int e^{-\beta V(\r)}\d\r
\ee
where $\lambda_\beta=\hbar\sqrt{2\pi\beta/m}$,
this holds if $V$ increases faster than logarithmically.)
As it was shown in \cite{S}, at any temperature the $N\to\infty$ limit of the thermal equilibrium state
of the corresponding noninteracting gas
remains essentially a finite perturbation of the ground state.
For this reason, at any finite temperature there is an asymptotically
complete condensation into $\varphi_0$, the ground state of $H^0$.
Because the ground state is common for the Bose and Boltzmann gases, there is
BEC, although incomplete, at any temperature even in the trapped ideal quantum
Boltzmann gas! The reduced one-particle density matrix of the latter
is $Ne^{-\beta H^0}/\tr e^{-\beta H^0}$, whose largest eigenvalue $Ne^{-\beta \veps_0}/\tr e^{-\beta H^0}$
gives the mean number of particles in $\varphi_0$
($\veps_0$ is the lowest eigenvalue of $H^0$).

Adding a scaled pair interaction (\ref{scale}) with property (\ref{bnalfn})
to the noninteracting gas while
keeping the trap fixed has a nontrivial effect, which can be expected by comparing
energies: The total free energy of the ideal gas is of the order of 1 (apart from the
trivial ground state energy $N\veps_0$ which could be chosen to be zero), while the total
interaction energy under condition (\ref{uN}) is of the order of $N$.
BEC survives such a perturbation at all temperatures.
Working directly at zero temperature, in three dimensions for $b_N=N^2$ and
$\alpha_N=N$ (which is Gross-Pitaevskii scaling, satisfying (\ref{uN})) Lieb and Seiringer were
able to prove a
complete BEC into the minimizer of the Gross-Pitaevskii energy functional \cite{LS}. This is very
different from $\varphi_0$, also showing the nontrivial effect of the interaction.

The strength of an integrable
interaction can be measured either by its integral or by its scattering length.
Whether or not a pair interaction $u_N$ whose integral is of order $1/N$ counts to be weak from
the other point of view, depends
on the space dimension. If $u_N(\x)=\alpha_N^2u(\alpha_N\x)$, the scattering length of $u_N$
is $1/\alpha_N$ times the scattering length of $u$. In three dimensions (\ref{uN}) imposes
$\alpha_N\sim N$, thus both the integral and the scattering length are of order $1/N$. In two
dimensions $\int\alpha_N^2u(\alpha_N\x)\d\x=\int u(\x)\d\x$, so $u_N$ has to be chosen in
a different form in order to comply with (\ref{uN}). In one dimension, with the
necessary choice $\alpha_N\sim 1/N$, we get a scattering length $\sim N$, therefore the two criteria
of strength contradict each other.
This last example is a mean-field type interaction while e.g.
$\alpha_N=N^2$ and $b_N=N$ represent the opposite limit
in one dimension.

Further insight is obtained if scaling of the interaction
is transformed into scaling of the potential and the temperature:
\bea\label{int_to_pot}
\sum_{i=1}^N[-\h2m\Delta_{\x_i}+V(\x_i)]+\sum_{1\leq i<j\leq N}\alpha_N^2u(\alpha_N(\x_i-\x_j))
\phantom{aaaaaaaaaaaaaaaaaaaaaaaa}\nonumber\\
=\alpha_N^2\left\{\sum_{i=1}^N[-\h2m\Delta_{\y_i}+\alpha_N^{-2}V(\alpha_N^{-1}\y_i)]
+\sum_{1\leq i<j\leq N}u(\y_i-\y_j)\right\}
\eea
where $\y_i=\alpha_N\x_i$. Because $\alpha_N^2$ multiplies the inverse temperature $\beta$, we obtain
a joint limit during which $N\to\infty$ and in three dimensions the temperature goes to zero and the
trap opens in such a way that the particle density tends to zero, while in one dimension
the temperature goes to infinity and the trap closes so that the density
diverges. The three dimensional example suggests that
whatever high the unscaled temperature, Gross-Pitaevskii scaling
$\alpha_N=N$ may reduce the thermal equilibrium
state to the ground state.
It is at least true that BEC remains complete at
positive temperatures (see \cite{Sei} for the Gross-Pitaevskii
limit and the discussion of Section 2 for the general case (\ref{scale})).

The analysis of the behaviour of a many-body system
in a scaled external field or in the presence of scaled interactions
comes about naturally in the study of trapped Bose gases.
Taking the limit $N\to\infty$ and scaling certain quantities during this limit
is an abstraction to obtain a mathematically clear-cut
answer, not different in its philosophy from the thermodynamic limit in homogenous systems.
The most frequently used scaling limit is related to the Gross-Pitaevskii
energy functional \cite{DGPS,LSY1} which describes the ground-state
properties of dilute Bose gases. As mentioned earlier, in a three-dimensional trap
this scaling is characterized by fixing $Nl_N$, where $l_N$ is the scattering length
of the pair interaction, see also \cite{LS}. Considering the scattering length as an effective
diameter of the particle, space filling is seen to decrease as $1/N^2$ in this limit.
Condition (\ref{scale})-(\ref{bnalfn})
is a generalization of the Gross-Pitaevskii scaling of the interaction. As a different example, scaling
of the external confining potential appears in the theoretical discussion of BEC in
elongated traps, realized experimentally more recently \cite{Goe,Bou}.  Various
aspects of a one-dimensional
behaviour can become manifest because a tight transverse trapping makes transversal degrees of
freedom freeze out. Theoretically this can be achieved by applying different scalings of the confining
potential in the transverse and longitudinal
directions. These studies reveal a rich structure of low-dimensional
regimes \cite{LSY,PSW,Ger}.
Note also that a theoretical work on one-dimensional bosons in a scaled
harmonic trap was published well before the first experimental realization \cite{KvD}.

The present paper is
motivated by two naturally arising questions. What happens with BEC in dense or strongly interacting
trapped gases? How does interaction modify BEC occurring with a phase transition in the ideal trapped
gas? To study these questions,
first (Section 2)
for a gas in a fixed trap we relax the condition (\ref{uN}) at the price of not
being able to prove BEC, only GBEC. We obtain the best result for one dimensional anharmonic and other
superharmonic traps, i.e.
potentials with a faster than quadratic increase. We can prove a complete GBEC at any finite temperature
in the presence of unscaled pair interactions, that is, the total interaction energy increasing as $N^2$.
This result holds also for attractive interactions, showing that the collapse of an attractive Bose gas
can be considered as a GBEC. A box is a superharmonic trap and thus, in principle, our findings may have
implications for the homogenous $\delta$-gas. However, we can prove GBEC only if $c/\rho\to 0$ as
$N\to\infty$. In other words, we just fail to prove it for the interacting gas. A reason of this failure
may be that probably there is no Bose condensation, ordinary or generalized, in
the ground state of the homogenous soft-$\delta$-gas. Although our method is too weak to prove BEC
in homogenous interacting gases, some nontrivial bounds can be obtained and are presented in Section 2
(equations (\ref{homJ}) and (\ref{hombound})).

In the case of a gas in a fixed trap our proof either yields BEC/GBEC for any
finite temperature or does not yield it at all.
In the second part (Section 3) of the paper we are interested in the stability
against interactions of BEC which occurs through a phase transition in the
noninteracting gas.
Because $T_c=\infty$ for fixed traps, it is natural to look for a finite $T_c$
by opening the trap together with $N\to\infty$.
During this limit the separation of energy levels and, in particular,
the gap above the ground state tend to zero, a situation
similar to that occurring in the homogenous gas. As a consequence, the
difficulties to include interactions start to resemble those appearing in
homogenous systems. We consider only the harmonic
potential, $V(x)=m\omega^2x^2/2$. In this case
the relevant dimensionless parameter is $a=\hbar\omega\beta$.
In three dimensions it was found \cite{DGPS} that the $N$-dependent
critical temperature of the noninteracting gas is given by
$k_BT_c(N)=\hbar\omega [N/\zeta(3)]^{1/3}$ which is equivalent to say that
the replacement of $a$ by $aN^{-1/3}$ gives rise to BEC at
$a_c=\zeta(3)^{1/3}\approx 1.063$. Although the one-particle density of states
is different, this scaling qualitatively
reproduces the properties of the homogenous gas.
Since the interactions for which we can prove the survival of
condensation are too weak in three
dimensions ($o(N^{1/3})$, negligible on the scale $N$ of the free
energy of the noninteracting gas),
in Section 3.1 we consider the analogous problem in one dimension.
Earlier Ketterle and van Druten \cite{KvD} found that in the
noninteracting gas the $N$-dependent
critical temperature was $k_BT_c(N)=\hbar\omega N/\ln(2N)$.
We start with a detailed study of the noninteracting system.
We replace $\omega$ by $\omega\gamma_N$ or
$a$ by $a\gamma_N$ where $\gamma_N$ tends to zero and discuss the different
possibilities.
Depending on the decay rate of $\gamma_N$, $T_c$ may be infinite or zero or may
have a finite positive value. If $\gamma_N=\ln N/N$, $T_c$ is finite positive
with $a_c=1$ (equivalently, $k_BT_c(N)=\hbar\omega N/\ln N$; although we note that $T_c(N)$
is ill-defined and $\gamma_N=\ln\alpha N/N$ leads to $a_c=1$ for any fixed $\alpha>0$) and
with BEC for $a>1$ and no BEC for $a\leq 1$.
On the other hand, $T_c=\infty$ if $\gamma_N/(\ln N/N)\to \infty$ and
$T_c=0$ if $\gamma_N/(\ln N/N)\to 0$. In particular, for $\gamma_N=1/N$ we find
$T_c=0$ and an extensive free energy and can, therefore, identify this scaling
limit with that of the homogenous system.
Since exact computations are possible, for the
critical scaling we can obtain the asymptotic distribution of $n_0$,
the number of particles in the ground state, and its mean value together with
finite-size corrections.
We find that for $a<1$ the limit
distribution of $n_0/N^a$ is exponential and $\langle n_0\rangle/N^a\to 1$.
For $a>1$ the condensate density $\langle n_0\rangle/N\to 1-a^{-1}$ and the
fluctuation of $n_0$ is huge, $(\ln N/N)(n_0-\langle n_0\rangle)$
is asymptotically distributed following Gumbel's
law. This holds true at the critical point $a=1$ as well where
$\langle n_0\rangle=N\ln\ln N/\ln N+O(N/\ln N)$. Not surprisingly, the free
energy is subextensive, $F^0_N=-(\pi^2/6\beta a)N/\ln N+o(N/\ln N)$
for any $a>0$, with no singularity at $a=1$.
A subtle difference compared with the three dimensional case, revealing
itself only through the details described above, is
that at all temperatures the 
gas forms a complete
generalized Bose-Einstein condensate with the participation of at most $\sim N/\ln N$
one-particle levels. We prove this in Section 3.2. We also show there that
for $a<1$ a large number of low-lying levels are equally occupied by $N^a$ particles,
and for $a>1$ particles not condensed into the one-particle ground state
are in a generalized condensate at its critical point and there is no fragmented condensation,
$\langle n_i\rangle=o(N)$ for each $i>0$.
In Section 3.3
we study the effect of interactions.
We can only prove that at least a complete GBEC occurs at all temperatures
provided that $\langle U_N\rangle_{\beta H^0_N}/N\ln N$
goes to zero as $N$ tends to infinity. Here $U_N$ is the $N$-particle interaction energy and
$\langle U_N\rangle_{\beta H^0_N}$ is its mean value taken in the thermal equilibrium state of the
noninteracting gas in the scaled trap.
Thus $\langle U_N\rangle_{\beta H^0_N}$ can be of the order of $N$ or of slightly higher
order, so that $F^0_N/\langle U_N\rangle_{\beta H^0_N}\to 0$, showing that
the effect of the interaction is indeed nontrivial.
For scaled pair interactions the analog of (\ref{bnalfn}) is the
somewhat weaker condition (stronger interaction)
$
b_N=o(\alpha_N\ln N/N)\ .
$
The mathematical analysis in Section 3 is conceptually not difficult but it is
ramified and lengthy. Readers uninterested in details can
go directly to the assertions of Theorems 3, 4 and 5.
The paper ends with a Summary.


\section{Generalized Bose condensation in traps}
\subsection{Preliminaries}

The Hamiltonian of $N$ noninteracting bosons is
\be
H^0_N=\sum_{i=1}^N[-\h2m\Delta_{\x_i}+V(\x_i)]= \oplus_{i=1}^N H^0\ .
\ee
We allow the
confining potential $V$ and thus the one-particle Hamiltonian $H^0$ to depend on $N$, and omit
to indicate the resulting $N$-dependence of the eigenvalues
$\veps_0<\veps_1\leq\veps_2\leq\cdots$ and
eigenvectors $\varphi_j$. The $N$-particle interaction energy
and the corresponding Hamiltonian will be denoted by $U_N$ and $H_N=H^0_N+U_N$, respectively.
$U_N$ is supposed to be bounded from below. 
As in
\cite{S}, for a positive integer $J$ and a positive number $\delta$ we define the modified Hamiltonians
$H^0(J,\delta)$, $H^0_N(J,\delta)$ and $H_N(J,\delta)$ by replacing $\veps_j$ with $\veps_j+\delta$ for
$j\leq J$ in the spectral resolution of $H^0$. For an operator $A$,
$\langle A\rangle_{\beta H}
=\Tr A\,e^{-\beta H}/\Tr e^{-\beta H}$ with the trace taken in the $N$-particle symmetric subspace.
In particular,
$\langle n_j\rangle_{\beta H}$ is the mean occupation number of $\varphi_j$.

The basic estimate we use for proving BEC or GBEC in the interacting gas is the following.

\begin{lemma}\label{ineq}
For any $\beta>0$
\bea\label{lemineq}
\sum_{j=0}^J\langle n_j\rangle_{\beta H_N}
&\geq&
\sum_{j=0}^J\langle n_j\rangle_{\beta H^0_N(J,\delta)}
       -\frac{1}{\delta}[\langle U_N\rangle_{\beta H^0_N}-\inf U_N]\nonumber\\
&\geq&
\langle n_0\rangle_{\beta H^0_N(J,\delta)}
       -\frac{1}{\delta}[\langle U_N\rangle_{\beta H^0_N}-\inf U_N]\ .
\eea
\end{lemma}
This bound is a consequence of the identity
\be
\sum_{j=0}^J\langle n_j\rangle_{\beta H^{(0)}_N(J,\delta)}=-\frac{\partial}
{\partial(\beta\delta)}\ln\Tr e^{-\beta H^{(0)}_N(J,\delta)}\ ,
\ee
the convexity of the logarithm of the trace as a function of $\beta\delta$ and
Bogoliubov's convexity inequality \cite{ZB}
\be
\ln\frac{\Tr e^{-\beta H_1}}{\Tr e^{-\beta H_2}}\geq -\beta\langle H_1-H_2\rangle_{\beta H_2}\ ,
\ee
cf. the first part of the proof of Theorem 4.2 in \cite{S}.

Lemma \ref{ineq} is useless in the case of hard-core interactions which yield $-\infty$ on
the right-hand side of
the inequality (\ref{lemineq}), but in the case of integrable interactions it leads to nontrivial
results. Suppose that $V$ is fixed, and
\be
U_N(\x_1,\ldots,\x_N)=\sum_{i<j}u_N(\x_i-\x_j)
\ee
One can show that
\be\label{diffU}
\frac{1}{N}|\langle U_N\rangle_{\beta H^0_N}-(\Phi_0,U_N\Phi_0)|\leq c(\beta)\|u_N\|_1
\ee
where $\Phi_0=\varphi_0(\x_1)\cdots\varphi_0(\x_N)$, $\|u_N\|_1=\int|u_N(\x)|\d\x$
and $c(\beta)$ is independent of $N$. The above lemma and the bound (\ref{diffU})
were used earlier to prove a theorem (Theorem 4.2 of Ref. \cite{S}) that we repeat here
in a slightly different form.

\begin{theorem}
\label{th1}
Suppose that $V$ is independent of $N$.
Let $U_N$ be a positive pair interaction with $\|u_N\|_1\leq C/N$ for some constant $C$. Then
\be
L(U)\equiv\lim_{N\to\infty}(\Phi_0,U_N\Phi_0)/N<\infty\ ,
\ee
and for any $\beta>0$ and $J\geq 0$
\be\label{BEC}
\sum_{j=0}^J\lim_{N\to\infty}\frac{1}{N}\langle n_j\rangle_{\beta H_N}\geq 1-\frac{L(U)}
{\veps_{J+1}-\veps_0}\ .
\ee
\end{theorem}
%
The main example is (\ref{scale}) with (\ref{bnalfn}).
Because $L(U)$ is finite and $\veps_j$ are independent of $N$ and tend to infinity,
there exists a $J_0$ independent of $N$ and uniquely defined through
\be
\veps_{J_0}-\veps_0\leq L(U)<\veps_{J_0+1}-\veps_0\ .
\ee
The right-hand side of
(\ref{BEC}) is positive for $J\geq J_0$, implying that
at least one of $\varphi_0\ldots,\varphi_{J_0}$ is macroscopically occupied.
This is just Bose-Einstein condensation, because of
the following simple result, shown in \cite{S}.
\begin{lemma}
Let $\sigma(N)$ be any $N$-particle density matrix and $\varphi$ any normalized element of the
one-particle Hilbert space ${\cal H}$. Then
\be
\langle n[\varphi]\rangle_\sigma=(\varphi,\sigma_1\varphi)
\ee
where $\sigma_1(N)$ is the one-particle reduced density matrix corresponding to $\sigma$
and $n[\varphi]$ is the occupation number operator in ${\cal H}^N$, associated to $\varphi$,
\be
n[\varphi]=|\varphi\rangle\langle\varphi|\otimes I\otimes\cdots\otimes I
+\cdots+I\otimes\cdots\otimes I\otimes|\varphi\rangle\langle\varphi|\ .
\ee
\end{lemma}
Thus
$(\varphi_j,\sigma_1\varphi_j)=\langle n[\varphi_j]\rangle_{\beta H_N}
\equiv\langle n_j\rangle_{\beta H_N}$ is of the order of $N$ for at least one of
$j=0,\ldots,J_0$, and the maximum eigenvalue of $\sigma_1$ satisfies
\be
\|\sigma_1\|\geq\frac{N}{J_0+1}\left(1-\frac{L(U)}{\veps_{J_0+1}-\veps_0}\right),
\ee
implying BEC.
Although in \cite{S} the theorem was proved for a general $J$ as presented above, it was
stated only for $J=J_0$, because we were interested in finding the minimum number of eigenstates
of $H^0$ one of which can be seen to be occupied macroscopically. In fact,
Theorem \ref{th1} implies a {\em complete} BEC at all temperatures. To see this,
it suffices to take the limit $J\to\infty$ in (\ref{BEC}), yielding
\be\label{suminf}
\sum_{j=0}^\infty\lim_{N\to\infty}\frac{1}{N}\langle n_j\rangle_{\beta H_N}=1.
\ee
Only $\langle n_j\rangle_{\beta H_N}\propto N$ contribute to the sum.
Thus, all but an asymptotically vanishing fraction of particles are carried by macroscopically occupied
one-particle eigenstates. It is not difficult to show that the same is true for the eigenstates
of $\sigma_1$. If $\sigma_1\psi_j=\lambda_j\psi_j$, $j=0,1\ldots$,
\be\label{sumlambda}
\sum_{j=0}^\infty\lim_{N\to\infty}\frac{1}{N}\langle n[\psi_j]\rangle_{\beta H_N}=
\sum_{j=0}^\infty\lim_{N\to\infty}\lambda_j(N)/N=1
\ee
comes from (\ref{BEC}) and the following generalization of the variational principle.

\begin{lemma}\label{vari}
Let $A$ be an upper semibounded self-adjoint operator on a separable Hilbert space ${\cal H}$. Suppose
that $A$ has a pure point spectrum $\lambda_1\geq\lambda_2\geq\ldots$. Then for any positive integer $n$
\be\label{eqlemma}
\sum_{i=1}^n\lambda_i=\sup_{\begin{array}{c}
                                     \phi_1,\ldots,\phi_n\in{\cal H}\\
				     (\phi_i,\phi_j)=\delta_{ij}
				     \end{array}
				    }
\sum_{i=1}^n(\phi_i,A\phi_i)\ .
\ee
\end{lemma}

\noindent
{\it Proof.}
Let $\psi_i$ be the orthonormal eigenvectors of $A$. Because $\lambda_i=(\psi_i,A\psi_i)$, the
left side of (\ref{eqlemma}) is smaller than or equal to the right side.
Therefore, only the opposite inequality is to be shown. Choose any orthonormal set $\phi_1,\ldots,\phi_n$
of vectors of ${\cal H}$ and let $\phi_i=\sum_{j=1}^\infty a_{ij}\psi_j$. Then
\bea
\sum_{i=1}^n(\phi_i,A\phi_i)
&=&\sum_{i=1}^n\sum_{j=1}^\infty |a_{ij}|^2\lambda_j
=\sum_{i=1}^n\left[\sum_{j=1}^n |a_{ij}|^2\lambda_j+\sum_{j=n+1}^\infty |a_{ij}|^2\lambda_j\right]
\nonumber\\
&\leq& \sum_{i=1}^n\left[\sum_{j=1}^n |a_{ij}|^2\lambda_j+\lambda_{n+1}
\left(1-\sum_{j=1}^n |a_{ij}|^2\right)\right]
=n\lambda_{n+1}+\sum_{i=1}^n\sum_{j=1}^n |a_{ij}|^2(\lambda_j-\lambda_{n+1})\nonumber\\
&=&n\lambda_{n+1}+\sum_{j=1}^n\left(\sum_{i=1}^n |a_{ij}|^2\right)(\lambda_j-\lambda_{n+1})
\leq n\lambda_{n+1}+\sum_{j=1}^n(\lambda_j-\lambda_{n+1})
=\sum_{j=1}^n\lambda_j\ ,
\eea
which remains valid if we take the supremum in the leftmost member.

\vspace{2mm}
Applying this lemma to $A=\sigma_1(N)$
we see that
$\sum_{j=0}^J\langle n_j\rangle_{\beta H_N}\leq\sum_{j=0}^J\lambda_j$ for any $J$. From (\ref{BEC}), therefore,
\be
\sum_{j=0}^J\lim_{N\to\infty}\lambda_j(N)/N\geq 1-\frac{L(U)}{\veps_{J+1}-\veps_0}\ .
\ee
Lettig $J$ tend to infinity we obtain (\ref{sumlambda}). For Gross-Pitaevskii scaling in three
dimensions Seiringer \cite{Sei} proved the much stronger result $\lim_{N\to\infty}\lambda_0(N)/N=1$.

The logic of the use of Lemma \ref{ineq} for proving GBEC
is somewhat different from the strategy followed in Theorem \ref{th1}:
First we make an appropriate choice of $J$ and then we impose the necessary
condition on the interaction. An obvious consequence of Lemma \ref{ineq} is

\begin{prop}\label{propgen}
Let $J=J(N)$ with $J=o(N)$ or $J=\lfloor sN\rfloor$. Choose
$\delta=\veps_{J+1}-\frac{1}{2}(\veps_0+\veps_1)$.
Fix $\beta>0$ and suppose that
\be\label{propcond}
\lim \frac{1}{N}\left[\sum_{j=0}^J\langle n_j\rangle_{\beta H^0_N(J,\delta)}
-\frac{1}{\delta}[\langle U_N\rangle_{\beta H^0_N}-\inf U_N]\right]=b>0
\ee
where $\lim$ means $\lim_{N\to\infty}$ if $J/N\to 0$ and $\lim_{s\to0}\lim_{N\to\infty}$ if
$J=\lfloor sN\rfloor$.
Then
\be\label{gen}
\lim\frac{1}{N}\sum_{j=0}^J\langle n_j\rangle_{\beta H_N}\geq b\ .
\ee
\end{prop}
The inequality (\ref{gen}) implies {\em at least} GBEC. Note that $\delta=\delta(J,N)$ with a separate
dependence on $N$ through the eigenvalues of $H^0$ if $V$ is scaled. In some applications
the positivity of the left member of (\ref{propcond}) can hold with a single term, $j=0$, of the sum.

The results of the theorem and the proposition above apply to the ground state as well, if we
take first the limit $\beta\to\infty$, then $N\to\infty$.
In the use of Lemma \ref{ineq} for proving BEC or GBEC
the lower part of the spectrum of $H^0$ is shifted upwards in such a way that
the ground state of the modified
Hamiltonian $H^0(J,\delta)$ remains $\varphi_0$, the ground state of $H^0$. Therefore
\be
\lim_{\beta\to\infty}\langle n_0\rangle_{\beta H^0_N(J,\delta)}=N
\ee
and there is BEC or GBEC in the ground state of the interacting gas provided that
\be\label{gsineq}
\lim \frac{1}{N\delta}[(\Phi_0,U_N\Phi_0)-\inf U_N]<1\ .
\ee

It is interesting to see why and how the method described above fails in proving BEC or GBEC of the
homogenous Bose gas. We exhibit this failure on the ground state. Let us consider first
the one-dimensional homogenous $\delta$-gas, that is, $u(x)=2c\delta(x)$ and
\bea\label{box}
V(x)=\left\{\begin{array}{cl}
0&0<x<L\\
+\infty&x\leq 0,\ x\geq L
\end{array}\right.
\eea
or take a periodic boundary condition at 0 and $L$.
Let $\rho=N/L$. In the periodic case
\(
(\Phi_0,U_N\Phi_0)=cN(N-1)/L
\)
and therefore GBEC follows from Proposition \ref{propgen} and equation (\ref{gsineq}) if $\lim J/N=0$ and
$c\rho<\delta\approx\veps_J\propto J^2/L^2=\rho^2(J/N)^2$
which means $\lim c/\rho=\lim (J/N)^2=0$, i.e. a vanishing interaction.
Still, Proposition \ref{propgen} yields a nontrivial estimate
for interacting homogenous gases. Let, in general,
$U_N=\sum_{i<j}u(\x_i-\x_j)\geq -BN$, $\int|u|\d\x<\infty$, $\rho>0$
and consider $N$ bosons in a $d$-dimensional cube of side $L=(N/\rho)^{1/d}$. For
periodic boundary conditions $(\Phi_0,U_N\Phi_0)=\frac{1}{2}\rho(N-1)\int u\d\x$ and therefore
(\ref{gsineq}) holds true if $\delta>\frac{1}{2}\rho\int u\d\x +B>0$.
Recalling that in the setup of the proposition $\lim_{N\to\infty}\veps_{J}/\delta=1$ if
$J\to\infty$ with $N$, choose
\be\label{homJ}
J(N)=[v_d(2m\delta)^{d/2}/\rho h^d]N
\ee
where $v_d$ is the volume of the unit ball in $d$ dimensions. Then the zero temperature version
of Proposition \ref{propgen} for homogenous gases proves
\be\label{hombound}
\lim_{N\to\infty}\lim_{\beta\to\infty}
\frac{1}{N}\sum_{j=0}^{J(N)}\langle n_j\rangle_{\beta H_N}\geq
1-\frac{\frac{1}{2}\rho\int u\d\x +B}{\delta}\ ,
\ee
i.e. a macroscopic number of particles are distributed over
a macroscopic number (\ref{homJ}) of lowest lying levels. Although this does not prove BEC,
the bound (\ref{hombound}) is nontrivial because the number of levels is
infinite for finite $N$.

To use Proposition \ref{propgen} at positive temperatures, BEC or GBEC has to be proven in the
noninteracting gas with the shifted spectrum.
This is a minor problem if the external potential is fixed, as in the case of our forthcoming
discussion of
GBEC in superharmonic traps. For the scaled harmonic potential we will have to pay somewhat
more attention to this question.

\subsection{Superharmonic traps in one dimension}

In one dimension the particularity of potentials with $x^2/V(x)\to 0$ as $|x|\to\infty$ is that
$n/\veps_n\to 0$ as $n\to\infty$. As an example, for homogenous potentials $V(x)=c|x|^\eta$
semiclassical quantization yields the eigenvalues in the form
\be\label{semiclass}
\veps_n=\left[\frac{h\eta c^{1/\eta}}{4\sqrt{2m} I_\eta}(n+O(1))\right]^{\frac{2\eta}{2+\eta}}
\qquad I_\eta=\int_0^1x^{-1+1/\eta}\sqrt{1-x}\d x
=\frac{\Gamma(\frac{3}{2})\Gamma(1/\eta)}{\Gamma(\frac{3}{2}+1/\eta)}\ ,
\ee
showing that the eigenvalues increase faster than linearly if $\eta>2$. Note that Sturmian theory
\cite{Tit} confirms the semiclassical formula (\ref{semiclass}).

Let, therefore, $H^0$ be a fixed one-particle Hamiltonian with a spectrum such that $n/\veps_n\to 0$
(which is our definition of a superharmonic trap in one dimension)
and choose $J(N)\to \infty$ in such a way that
$\lim J/N=0$ and $\lim N/\veps_J=0$. Then for
$\delta=\veps_{J+1}-\frac{1}{2}(\veps_0+\veps_1)$ we also have
$\lim N/\delta=0$.

In \cite{S} we proved that in the case of a noninteracting gas in a fixed trap
there exists a uniform upper bound on the mean value
of $N'=N-n_0$, namely, for any $N$ and any $\mu<\veps_1-\veps_0$
\be
\langle N'\rangle_{\beta H^0_N}\leq\frac
{1}{(1-e^{-\beta\mu})^2}\prod_{n=1}^\infty
\frac{1}{1-e^{-\beta(\veps_n-\veps_0-\mu)}}\ .
\ee
Similar inequality holds for $\langle N'\rangle_{\beta H^0_N(J,\delta)}$,
with the exception that the bound still depends on $N$ because $\veps_n-\veps_0$ has to be
replaced by $\veps_n-\veps_0-\delta$ if $n>J$. However, due to superharmonicity, for $n\geq J+2$
we can use the estimate
\be
\veps_n-\veps_0-\delta=\veps_n-\veps_{J+1}+\frac{1}{2}(\veps_1-\veps_0)>\veps_{n-J-1}-\veps_0
\ee
if $N$ (and thus $J$) is large enough. Choosing e.g. $\mu=\frac{1}{4}(\veps_1-\veps_0)$,
for sufficiently large $N$ we obtain
\be
\langle N'\rangle_{\beta H^0_N(J,\delta)}\leq\frac
{1}{(1-e^{-\beta\mu})^3}\left[\prod_{n=1}^\infty
\frac{1}{1-e^{-\beta(\veps_n-\veps_0-\mu)}}\right]^2\ .
\ee
Therefore
\be
\lim\frac{1}{N}\langle n_0\rangle_{\beta H^0_N(J,\delta)}=1
\ee
and the inequality (\ref{propcond}) holds if
\be\label{undelta}
\lim\frac{1}{N\delta}[\langle U_N\rangle_{\beta H^0_N}-\inf U_N]<1\ .
\ee
In words, GBEC follows if the mean interaction energy
$\langle U_N\rangle_{\beta H^0_N}$ does not exceed $N\delta\approx N\veps_J$, the maximum
energy of $N$ noninteracting particles in the would-be (generalized) condensate.

If $U_N$ is a pair interaction, we can use the estimate (\ref{diffU}) and
\be
|(\Phi_0,U_N\Phi_0)|\leq\frac{N(N-1)}{2}\|u_N\|_1 \|\varphi_0^4\|_1
\ee
to obtain
\bea\label{limndel}
\lim\frac{1}{N\delta}[\langle U_N\rangle_{\beta H^0_N}-\inf U_N]
\leq\lim\frac{1}{\delta}\left[\left(\frac{N-1}{2}\|\varphi_0^4\|_1+c(\beta)\right)\|u_N\|_1-
\frac{\inf U_N}{N}\right].
\eea
If the interaction is stable, $\inf U_N/N\geq -B$ for some constant $B$; if it is not stable still
$\inf U_N/N\geq -(N-1)\inf u_N/2$.
In either case, for $u_N=u$ integrable, bounded from below and independent of $N$ the quantity in the square
bracket is of the order of $N$ and thus the limit in (\ref{limndel}) vanishes. With Lemma \ref{vari}
this implies

\begin{theorem}\label{superharm}

In one-dimensional superharmonic traps bosons interacting via a lower semibounded (unscaled)
integrable pair interaction
undergo a complete generalized Bose-Einstein condensation at all temperatures:
For any choice of $J(N)$ such that
$\lim J/N=0$ and $\lim N/\veps_J=0$
\be
\lim\frac{1}{N}\sum_{j=0}^J\lambda_j(N)=
\lim\frac{1}{N}\sum_{j=0}^J\langle n_j\rangle_{\beta H_N}=1\ .
\ee

\end{theorem}

The claim of the theorem could be made stronger. For instance,
in a box (\ref{box}) with a fixed $L$ independent of $N$, $J$ and $\delta\sim\veps_J$
can grow almost as fast as $N$ and $N^2$, respectively, and therefore
we can obtain GBEC for interactions as strong as $\langle U_N\rangle=o(N^3)$ instead of $O(N^2)$ stated
in the theorem.

\section{Scaled harmonic trap in one dimension}
\subsection{Bose condensation via phase transition in the noninteracting gas}

The one-particle Hamiltonian is
\be
H^0=-\h2m \frac{\d^2}{\d x^2}+\frac{1}{2}m\omega^2 x^2\ .
\ee
Measuring the energy from that of the ground state, the partition function for $N$ noninteracting
particles reads
\be\label{Qq}
Q_{N,a}=\sum_{\{n_j\}_{j\geq 0}:\sum n_j=N}e^{-a\sum jn_j}
=\sum_{\{n_j\}_{j>0}:\sum n_j\leq N}e^{-a\sum jn_j}
=\sum_{m=0}^N q_{m,a}
\ee
with
\be
q_{m,a}=\sum_{\{n_j\}_{j>0}:\sum n_j=m}e^{-a\sum jn_j}
\ee
where we have introduced the notation $a=\hbar\omega\beta$ and
used $\beta(\veps_j-\veps_0)=ja$.
The key to the forthcoming analysis is

\begin{lemma}\label{QqPna}
\be\label{Q}
Q_{N,a}=\prod_{k=1}^N(1-e^{-ka})^{-1}
\ee
\be\label{q}
q_{m,a}=e^{-ma}\prod_{k=1}^m(1-e^{-ka})^{-1}
\ee
and the probability of having $m$ particles in excited states is
\be\label{pnaeqm}
\pna(N'=m)=e^{-ma}\prod_{k=m+1}^N(1-e^{-ka})\ .
\ee
\end{lemma}
{\em Proof.}\\
$\pna(N'=m)=q_{m,a}/Q_{N,a}$ and 
(\ref{q}) follows from (\ref{Q}) through $q_{m,a}=Q_{m,a}-Q_{m-1,a}$.
Now $Q_{N,a}$ can be rewritten as
\bea
Q_{N,a}
=\sum_{0\leq i_1\leq\cdots\leq i_N}e^{-a\sum_{j=1}^Ni_j}
=\sum_{i_1=0}^\infty e^{-i_1a}\sum_{i_2=i_1}^\infty e^{-i_2a}\cdots\sum_{i_N=i_{N-1}}^\infty e^{-i_Na}
\eea
from which (\ref{Q}) follows.

An alternative way is to compute first $q_{m,a}$ by using that it is the
generating function of $p_m(n)$, the number of (unordered) partitions of $n$ into $m$ parts,
\be
q_{m,a}=\sum_{n=m}^\infty p_m(n)e^{-na}=\sum_{n=1}^\infty p_m(n)e^{-na}
\ee
because $p_m(n)=0$ for $n<m$. Starting with the identity
\be
p_m(n)=p_m(n-m)+p_{m-1}(n-m)+\cdots+p_1(n-m)\ ,
\ee
valid for $n>m\geq 1$, one can derive the recurrence relation
\be\label{qQ}
q_{m,a}=\frac{e^{-ma}}{1-e^{-ma}}\sum_{k=0}^{m-1}q_{k,a}
\qquad q_{0,a}\equiv 1
\ee
from which (\ref{q}) follows by induction and (\ref{Q}) by (\ref{Qq})  and (\ref{qQ}).

\vspace{2mm}
Due to the simple form of $\pna(N'=m)$ we can obtain precise asymptotic results on the distribution
of $n_0$ in the case of different scalings. In what follows, we discuss the thermodynamics of the
noninteracting gas in a scaled harmonic trap, characterized by the Hamiltonian
\be\label{harmonic}
H^0_N=\sum_{i=1}^N[-\h2m\frac{\d^2}{\d x_i^2}+\gamma_NV(\sqrt{\gamma_N}x_i)]
\ee
where $V(x)=\frac{1}{2}m\omega^2x^2$ and $\gamma_N\to 0$. This amounts to open the trap by replacing
$\omega$ by $\omega\gamma_N$ or, equivalently, $a=\hbar\omega\beta$ by
\be\label{aNgen}
a_N=a\gamma_N\to 0.
\ee
From Lemma \ref{QqPna}
\be\label{N'leqm}
\pnan(N'\leq m)=\frac{Q_{m,a_N}}{Q_{N,a_N}}=\prod_{l=m+1}^N(1-e^{-la_N})\ .
\ee

Let $\lambda_N=1-m/N$ where $m$ is any integer between 0 and $N$.
Taking the logarithm of equation (\ref{N'leqm}) and expanding the right member in Taylor series we find
\be\label{logprob}
\ln\pnanl=-\sum_{k=1}^\infty A_{Nk}\ ,\quad
A_{Nk}(\lambda_N)=\frac{1}{k}\sum_{l=m+1}^Ne^{-kla_N}
=\frac{e^{-kNa_N}(e^{k\lambda_NNa_N}-1)}{k(e^{ka_N}-1)}\ .
\ee
$A_{Nk}(\lambda_N)$ are nonnegative,
monotone decreasing with $k$ and increasing with $\lambda_N$. Later on,
\be\label{Anineq}
\pnanl\leq e^{-A_{N1}(\lambda_N)}
\ee
will be used.

The following two propositions serve to find necessary conditions on the sequence $a_N$ which give rise to
a trivial asymptotic distribution of $n_0/N$.
In this section we use the notation $\langle n_0\rangle_{N,a_N}$ for
$\langle n_0\rangle_{\beta H^0_N}$.

\begin{prop}\label{3.2}
Suppose that
\be\label{cond3.1}
\lambda_N\to\lambda>0\quad{\rm and}\quad\ln N-\left(1-\frac{\lambda_N}{2}\right)Na_N\to\infty\ .
\ee
Then
\be\label{double}
\pnanl\to 0\quad {\rm and}\quad
\lim_{N\to\infty}\frac{1}{N}\langle n_0\rangle_{N,a_N}\leq\lambda\ .
\ee
If $\ln N-Na_N\to\infty$ then $\frac{1}{N}\langle n_0\rangle_{N,a_N}\to 0$,
i.e. there is no Bose-Einstein condensation.
\end{prop}
{\em Proof.}\\
To obtain the vanishing probability it suffices to show that $A_{N1}\to\infty$.
From the inequalities
\be\label{xineq}
xe^{x/2}<e^x-1<xe^x
\ee
valid for $x>0$,
\be
A_{N1}>e^{\ln N-(1-\lambda_N/2)Na_N+\ln\lambda_N-a_N}\to\infty
\ee
indeed. Moreover,
\be\label{nnul<}
\frac{1}{N}\langle n_0\rangle_{N,a_N}\leq\lambda_N\pnan\left(\frac{n_0}{N}<\lambda_N\right)
+\pnanl\to \lambda.
\ee
If the stronger condition $\ln N-Na_N\to\infty$ is fulfilled, (\ref{cond3.1}) and
therefore (\ref{double}) hold true
for all $\lambda>0$.  This implies the absence of BEC.

\vspace{2mm}
The counterpart of Proposition \ref{3.2} is

\begin{prop}\label{3.3}
Suppose that
\be\label{cond3.3}
\lambda_N\to\lambda\leq 1\quad{\rm and}\quad(1-\lambda_N)Na_N-\ln N\to\infty\ .
\ee
Then
\be\label{double1}
\pnanl\to 1\quad {\rm and}\quad
\lim_{N\to\infty}\frac{1}{N}\langle n_0\rangle_{N,a_N}\geq\lambda\ .
\ee
If $Na_N/\ln N\to\infty$ then $\frac{1}{N}\langle n_0\rangle_{N,a_N}\to 1$,
i.e. there is a complete Bose-Einstein condensation.
\end{prop}
{\em Proof.}\\
Using (\ref{xineq}) and
$$-\ln(1-e^{-x})<\frac{1}{e^x-1}\qquad (x>0)$$
we obtain
\be
0<\sum_{k=1}^\infty A_{Nk}<N\lambda_N\left|\ln\left(1-e^{-(1-\lambda_N)Na_N}\right)\right|
<\frac{\lambda_N}{e^{(1-\lambda_N)Na_N-\ln N}-\frac{1}{N}}\to 0
\ee
which implies the result on the limit of the probability. On the other hand,
\be\label{full}
\frac{1}{N}\langle n_0\rangle_{N,a_N}\geq\lambda_N\pnanl\to \lambda\ .
\ee
Suppose now that $Na_N/\ln N\to\infty$. One can choose a sequence $\lambda_N\to 1$
in such a way that $(1-\lambda_N)Na_N/\ln N\to\infty$. Then (\ref{cond3.3})
is fulfilled and (\ref{full}) holds for $\lambda=1$.

\vspace{2mm}
From Propositions \ref{3.2} and \ref{3.3} we can identify the scalings which yield trivial limits:

\begin{coro}\label{cortrivi}
If $N\gamma_N/\ln N\to 0$ then $T_c=0$. If $N\gamma_N/\ln N\to\infty$ then $T_c=\infty$ and there is a
complete BEC at all $T<\infty$.
\end{coro}
{\em Proof.}\\
In the first case $Na_N/\ln N\to 0$ and thus $\ln N-Na_N\to \infty$ for any $a<\infty$ i.e. $T>0$, and
Proposition \ref{3.2} applies.
In the second case $Na_N/\ln N\to\infty$ for any $a>0$ i.e. $T<\infty$, and Proposition \ref{3.3} applies.

\vspace{2mm}
Thus, we have found that a phase transition can occur only if $N\gamma_N/\ln N$ has a finite nonvanishing
limit. In this case without restricting generality we suppose that
\be\label{aNspec}
\gamma_N=\ln N/N, \quad a_N=a\ln N/N.
\ee
Choosing a different prefactor can only change the critical temperature. Propositions \ref{3.2}
and \ref{3.3} already yield the phase transition:

\begin{coro}
If $a_N=a\ln N/N$ then $a_c=1$. For $a<1$ there is no BEC. For $a>1$ and
$0<\lim\lambda_N<1-a^{-1}$
\be
\lim_{N\to\infty}\pnanl=1.
\ee
Moreover,
\be\label{coro2}
\lim_{N\to\infty}\frac{1}{N}\langle n_0\rangle_{N,a_N}\geq 1-\frac{1}{a}.
\ee
\end{coro}
{\em Proof.}\\
If $a<1$, $\ln N-Na_N=(1-a)\ln N\to\infty$ and by Proposition \ref{3.2} there is no BEC. If $a>1$ and
$0<\lambda=\lim\lambda_N<1-a^{-1}$ then (\ref{cond3.3}) and thus (\ref{double1}) hold true.
Letting $\lambda$ tend to
$1-a^{-1}$ we find (\ref{coro2}).

\vspace{2mm}
This is a temporary result. We will show, among others, that for $a\geq 1$
the probability of having $n_0/N>1-a^{-1}$ tends to zero and thus the distribution of $n_0/N$ becomes
degenerate and concentrated on $1-a^{-1}$. This will imply that in (\ref{coro2}) there is equality.

\begin{prop}\label{prop3.3}
Suppose that
\be\label{cond3.4}
\lambda_N\to\lambda\leq 1\quad{\rm and}\quad(1-\lambda_N)Na_N\to\infty\ .
\ee
Then
\be\label{res3.4}
\pnanl=e^{-(1+\epsilon_N)A_{N1}(\lambda_N)}
\ee
where
$0<\epsilon_N<-\ln\left(1-e^{-(1-\lambda_N)Na_N}\right)\to 0$.
\end{prop}
{\em Proof.}\\
Let us rewrite equation (\ref{logprob}) in the form
\be
\ln\pnanl=-A_{N1}\left(1+\sum_{k=1}^\infty\frac{A_{N,k+1}}{A_{N1}}\right).
\ee
Now
\bea
\frac{A_{N,k+1}}{A_{N1}}&=&\frac{e^{-(m+1)(k+1)a_N}}{e^{-(m+1)a_N}}\frac{1}{k+1}
\frac{\sum_{l=1}^{N-m-1}e^{-{l(k+1)a_N}}}{\sum_{l=1}^{N-m-1}e^{-{la_N}}}\nonumber\\
&<&\frac{1}{k+1}
e^{-(m+1)ka_N}<\frac{1}{k}e^{-k(1-\lambda_N)Na_N}
\eea
and therefore
\be\label{domin}
\sum_{k=1}^\infty\frac{A_{N,k+1}}{A_{N1}}<\sum_{k=1}^\infty
\frac{1}{k}e^{-k(1-\lambda_N)Na_N}
=-\ln\left(1-e^{-(1-\lambda_N)Na_N}\right)\to 0
\ee
as $N\to\infty$.

\vspace{2mm}
Henceforth, we concentrate on the phase transition.
With the scaling (\ref{aNspec}) the condition (\ref{cond3.4}) is satisfied
if $a>0$ and $\lambda<1$.
Most of the results listed in the theorem below follow from
Proposition \ref{prop3.3} applied to this case,
\be\label{specasym}
\pnanl=\exp\left\{-N^{-a}\frac{N^{a\lambda_N}-1}{e^{a\ln N/N}-1}(1+\epsilon_N)\right\}.
\ee

\begin{theorem}\label{BECharmfree}
The scaling $a_N=a\ln N/N$ leads to a phase transition at
$a\equiv \hbar\omega\beta=1$ with no
BEC for $a\leq 1$ and BEC for $a>1$.
In details, the following hold true.

\noindent
I. Limit distribution of $n_0$.

(i) For $0<a<1$ and $x\geq 0$
\be\label{exp}
\lim_{N\to\infty}\pnan\left(\frac{n_0}{N^a}\geq x\right)=e^{-x}.
\ee

(ii) For $a\geq 1$ and $\lambda_N\to\lambda\in ]0,1[$
\bea\label{step}
\lim_{N\to\infty}\pnanl=\left\{\begin{array}{cll}
                                   1&{\rm if}&\lambda<1-a^{-1}\\
				   0&{\rm if}&\lambda>1-a^{-1}.
				   \end{array}\right.
\eea

(iii) For $a\geq 1$ and any real $x$
\be\label{Gumbel}
\lim_{N\to\infty}\pnan\left(\left[\frac{n_0}{N}-1+a^{-1}\right]\ln N
-a^{-1}\ln\ln N\geq x\right)
=\exp\left\{-a^{-1}e^{ax}\right\}.
\ee
\phantom{aaaaa} Equivalently,
\be\label{Gumbel2}
\lim_{N\to\infty}\pnan\left(\frac{\ln N}{N}(n_0-\langle n_0\rangle_{N,a_N})
\geq x\right)=\exp\left\{-a^{-1}e^{a(x+\eta(a))}\right\}
\ee
\phantom{aaaaa} where
\be
\eta(a)=\int_{-\infty}^\infty x\exp\left\{ax-a^{-1}e^{ax}\right\}\d x\ .
\ee

\noindent
II. Mean value of $n_0$.

(i) For $0<a<1$
\be\label{meann0a<1}
\lim_{N\to\infty}\frac{\langle n_0\rangle_{N,a_N}}{N^a}=1.
\ee

(ii) For $a\geq 1$
\be\label{meandetailed}
\langle n_0\rangle_{N,a_N}= (1-a^{-1})N+\frac{N}{\ln N}[a^{-1}\ln\ln N
+\eta(a)]+o\left(\frac{N}{\ln N}\right).
\ee
\phantom{aaaaa} In particular,
\be\label{mean}
\lim_{N\to\infty}\frac{1}{N}\langle n_0\rangle_{N,a_N}= 1-\frac{1}{a}.
\ee

\noindent
III. For any $a>0$ the free energy of $N$ particles is
\be\label{free}
F^0_N=-\frac{\pi^2}{6\beta a}\frac{N}{\ln N}+o\left(\frac{N}{\ln N}\right).
\ee
\end{theorem}
{\em Proof.}

\noindent
(I.i) Substitute $\lambda_N=x/N^{1-a}$ into equation (\ref{specasym}).
Because $\lambda_N\ln N\to 0$, $N^{a\lambda_N}-1$ can be replaced by
$a\lambda_N\ln N$ and (\ref{exp}) follows.

\noindent
(I.ii) If $\lambda<1-a^{-1}$ then $A_{N1}(\lambda_N)\to 0$ and if $\lambda>1-a^{-1}$ then $A_{N1}(\lambda_N)\to\infty$.
Therefore the limit of (\ref{specasym}) is the degenerate distribution (\ref{step}).

\noindent
(I.iii) The Gumbel distribution (\ref{Gumbel}) is obtained by substituting
\be\label{lambdaGumbel}
\lambda_N=1-\frac{1}{a}+\frac{\ln\ln N+ax}{a\ln N}
\ee
into equation (\ref{specasym}). Because $\lambda_N\to 1-a^{-1}<1$, condition (\ref{cond3.4}) is
satisfied and Proposition \ref{prop3.3} indeed applies.
The form (\ref{Gumbel2}) follows from (II.ii) to be shown below; $\eta(a)$ is the expectation value
of the Gumbel distribution.

\noindent
(II.i) For any $\Delta>0$
\bea\label{sandwich}
\sum_{m=1}^\infty m\Delta\left[\pnan\left(\frac{n_0}{N^a}\geq m\Delta\right)-
\pnan\left(\frac{n_0}{N^a}\geq (m+1)\Delta\right)\right]\phantom{aaaaaaaaaaaaaaaa}\nonumber\\
\leq\frac{\langle n_0\rangle_{N,a_N}}{N^a}\leq
\sum_{m=1}^\infty m\Delta\left[\pnan\left(\frac{n_0}{N^a}\geq (m-1)\Delta\right)-
\pnan\left(\frac{n_0}{N^a}\geq m\Delta\right)\right]\ .
\eea
The sums are actually finite but the upper bounds tend to infinity with $N$.
Using the inequalities (\ref{Anineq}) and (\ref{xineq})
\be
\pnan\left(\frac{n_0}{N^a}\geq x\right)<\exp\{-A_{N1}(xN^{-1+a})\}<
\exp\{-xe^{-a\ln N/N}\}<e^{-x/2}
\ee
for $N$ large enough. For the left and right members of (\ref{sandwich}) $m\Delta e^{-(m-1)\Delta/2}$
is a summable upper bound, thus we can interchange the limit $N\to\infty$ and
the summation over $m$ to find, with (\ref{exp}) and the convexity of the
exponential,
\bea
e^{-\Delta}\sum_{m=1}^\infty m\Delta e^{-m\Delta}\Delta\leq
\sum_{m=1}^\infty m\Delta (e^{-m\Delta}-e^{-(m+1)\Delta})\leq
\lim_{N\to\infty}\frac{\langle n_0\rangle_{N,a_N}}{N^a}\nonumber\\
\leq
\sum_{m=1}^\infty m\Delta(e^{-(m-1)\Delta}-e^{-m\Delta})
\leq e^\Delta
\sum_{m=1}^\infty m\Delta e^{-m\Delta}\Delta\ .
\eea
Letting $\Delta$ tend to zero the Riemann sums go to
$\int_0^\infty xe^{-x}\d x=1$.

\noindent
(II.ii)
We start as before. Let
\be
f_N(n_0)=\left(\frac{n_0}{N}-1+a^{-1}\right)\ln N-a^{-1}\ln\ln N\ .
\ee
For any $\Delta>0$
\bea\label{sandwich2}
\sum_{m=-\infty}^\infty m\Delta\left[\pnan\left(f_N(n_0)\geq m\Delta\right)-
\pnan\left(f_N(n_0)\geq (m+1)\Delta\right)\right]\phantom{aaaaaaaaaaaaaaaaaaaaaa}
\nonumber\\
\leq\langle f_N(n_0)\rangle_{N,a_N}
\leq
\sum_{m=-\infty}^\infty m\Delta\left[\pnan\left(f_N(n_0)\geq (m-1)\Delta\right)-
\pnan\left(f_N(n_0)\geq m\Delta\right)\right]\ .
\eea
The sums are finite with the upper and lower bounds tending to infinity as $N$
increases.
Suppose that we can interchange the summation with the limit $N\to\infty$.
Then (\ref{Gumbel}) yields
\bea
\sum_{m=-\infty}^\infty m\Delta\left[e^{-a^{-1}e^{am\Delta}}
-e^{-a^{-1}e^{a(m+1)\Delta}}\right]\phantom{aaaaaaaaaaaaaaaaaaaaaaaaaaa}
\nonumber\\
\leq
\lim_{N\to\infty}\langle f_N(n_0)\rangle_{N,a_N} \leq
\sum_{m=-\infty}^\infty m\Delta\left[e^{-a^{-1}e^{a(m-1)\Delta}}
-e^{-a^{-1}e^{am\Delta}}\right]\ .
\eea
Now $\exp\{-a^{-1}e^{ax}\}$ is concave if $x<a^{-1}\ln a$ and convex if
$x>a^{-1}\ln a$. Accordingly, we divide the sums in two parts, bound the
differences in the square brackets with $\Delta$ times the derivatives
at the upper or lower end of the intervals and let $\Delta$ go to zero.
Both the upper and lower bound
of $\langle f_N(n_0)\rangle_{N,a_N}$
converge to $\eta(a)$.
Equation (\ref{meandetailed}) follows by simple rearrangement.
In (\ref{sandwich2})
the interchange of the summation with the limit $N\to\infty$ is again based on
the dominated convergence theorem. However, the sums have to be divided in
two parts. For $m>0$ we can use (\ref{Anineq}) with (\ref{lambdaGumbel}) and
$x=(m-1)\Delta$ as an upper bound on the difference (actually on both terms) in the square brackets
because $A_{N1}(\lambda_N)\geq (2a)^{-1}e^{ax}$ if $N$ is large enough.
For $m<0$ only the difference in the square bracket is small.
Using (\ref{domin}) a lengthy but straightforward computation yields
\bea
\pnan\left(f_N(n_0)\geq m\Delta\right)-
\pnan\left(f_N(n_0)\geq (m+1)\Delta\right) \leq
1-\exp\left\{-2a^{-1}(e^{a\Delta}-1)e^{am\Delta}\right\}
\eea
which decays exponentially as $m\to -\infty$ thus yielding a summable
upper bound.

Equation
(\ref{mean}) is a consequence of (\ref{meandetailed}) but can also be obtained
directly from the degenerate distribution (\ref{step}):
Choose any $\epsilon>0$.
\bea
\left(1-a^{-1}-\epsilon\right)\pnan\left(\frac{n_0}{N}
\geq 1-a^{-1}-\epsilon\right)
\leq\frac{\langle n_0\rangle_{N,a_N}}{N}\phantom{aaaaaaaaaaaaaaaaaaaaaaaaaaaa}
\nonumber\\
\leq
\left(1-a^{-1}+\epsilon\right)\pnan\left(\frac{n_0}{N}
\leq1-a^{-1}+\epsilon\right)
+\pnan\left(\frac{n_0}{N}>1-a^{-1}+\epsilon\right)\ .
\eea
Taking the limit $N\to\infty$ and applying (\ref{step}),
\be
1-a^{-1}-\epsilon
\leq\liminf\frac{\langle n_0\rangle_{N,a_N}}{N}\leq\limsup\frac{\langle n_0\rangle_{N,a_N}}{N}
\leq1-a^{-1}+\epsilon
\ee
which holds for any $\epsilon>0$ and thus for $\epsilon=0$ as well.

\noindent
III. If $Z_{N,a_N}$ denotes the $N$-particle partition function then
\be
F^0_N=-\frac{1}{\beta}\ln Z_{N,a_N}=-\frac{1}{\beta}
\left(-\frac{1}{2}Na_N+\ln Q_{N,a_N}\right).
\ee
From (\ref{Q})
\be
\ln Q_{N,a_N}=-\sum_{k=1}^N \ln(1-e^{-ka_N})
\ee
so that
\be\label{frenergy}
\lim_{N\to\infty}\gamma_N \ln Q_{N,a_N}=-\lim_{N\to\infty}\sum_{k=1}^N \gamma_N\ln(1-e^{-ak\gamma_N})
=-\int_0^\infty\ln(1-e^{-ax})\d x
\ee
because $\gamma_N\to 0$ and $N\gamma_N=\ln N\to\infty$. Expanding the logarithm and integrating term by
term ($\sum_{k=1}^n e^{-kax}/k\to-\ln(1-e^{-ax})$ monotonically)
we find $\pi^2/6a$ and, hence, equation (\ref{free}).

\vspace{2mm}
\noindent
{\it Remarks.}\\
(1) Equation (\ref{Gumbel2}) implies
$|n_0-\langle n_0\rangle_{N,a_N}|\sim N/\ln N$ for $a>1$. These are huge fluctuations even
compared with the super-normal fluctuations $|n_0-\langle n_0\rangle|\sim N^{2/3}$ in the condensation
regime of the three-dimensional homogenous Bose gas \cite{BP}.\\
(2) There is no singularity in
$
\lim\gamma_NF^0_N=-\pi^2/6\beta a
$
at the critical point $a=1$.\\
(3) For $\gamma_N=1/N$ equation (\ref{frenergy}) yields an extensive free energy
\be
\lim_{N\to\infty}\frac{1}{N}F^0_N
=-\beta^{-1}\lim_{N\to\infty}\frac{1}{N} \ln Q_{N,a_N}
=\beta^{-1}\int_0^1\ln(1-e^{-ax})\d x
\ee
so that this scaling corresponds to the homogenous limit. According to
Corollary \ref{cortrivi}, $T_c=0$ in this case as it has to be in the
one dimensional homogenous Bose gas.

\subsection{Generalized Bose condensation at all temperatures in the noninteracting gas}

In the three-dimensional homogenous noninteracting Bose gas above the critical temperature the
mean occupation number of each one-particle state remains finite in the thermodynamic limit.
This is immediately
seen in the grand-canonical ensemble and is valid in the canonical ensemble as well, due to the strong
equivalence of ensembles (the Kac density is a Dirac delta). The situation is quite different in the
one-dimensional
scaled harmonic trap when $\gamma_N=\ln N/N$. As we shall see, there is a complete GBEC at all temperatures.
It will also be shown that below the critical temperature the condensate is not fragmented,
$\langle n_i\rangle_{N,a_N}=o(N)$ for $i>0$. This means that for $a>1$ a condensate in the ground
state of $H^0$ whose density is $1-1/a$ coexists with a generalized condensate of density $1/a$.

The intuition behind the results of this section is guided by the following observation.
(We drop the subscript $a_N$ which plays no role here.)

\begin{lemma}
For any $i$
\be\label{ni-n0}
\langle n_i\rangle_{N}=\sum_{M=0}^N \langle n_0\rangle_{N-M}\
P_{N}\left(\sum_{j=0}^{i-1}n_j=M\right).
\ee
\end{lemma}

\noindent
{\it Proof.} The reader can easily check that because of
\[
\veps_n=\veps_{0}+n(\veps_1-\veps_0)
\]
the conditional distribution of $n_i$, given the number of particles in the lower lying eigenstates,
satisfies
\be
P_{N}\left(n_i=m\left|\sum_{j=0}^{i-1}n_j=M\right)=P_{N-M}(n_0=m)\right..
\ee
Multiplying by $m$ and summing over it the right member becomes $\langle n_0\rangle_{N-M}$ while on the
left-hand side we obtain the conditional expectation value $\langle n_i|n_0+\cdots+n_{i-1}=M\rangle_{N}$.
Multiplying by the probability of the condition and summing over $M$ yields (\ref{ni-n0}).

\vspace{2mm}
Because for $a<1$ $n_0\approx N^a$, we expect that
\be\label{nia<1}
\langle n_i\rangle_{N,a_N}\asymp \langle n_0\rangle_{N-iN^a,a_N}\asymp N^a
\ee
if $i\ll N^{1-a}$, and that an $o(N)$ number
of lowest lying levels carry roughly all the particles.
This should hold true for $a>1$ as well and, because of $n_0\approx N(1-1/a)$, we expect also that
\be\label{nia>1}
\langle n_i\rangle_{N,a_N}\asymp \langle n_0\rangle_{N/a-(i-1)a^{-1}N\ln\ln N/\ln N,a_N}\asymp a^{-1}
N\ln\ln N/\ln N
\ee
for $1\leq i\ll\ln N/\ln\ln N$.
Not all these conjectures will be verified below.
We start by proving the second part of (\ref{nia<1}).

\begin{prop}\label{propGBEC}
All the results of Theorem \ref{BECharmfree} remain valid if the scaling $a_N=a\ln N/N$ is
replaced by $a_N'=(1+\eta_N)a_N$ where $\eta_N=o(1/\ln N)$.
\end{prop}

\noindent
{\em Proof.} Consider $A_{Nk}$, defined in (\ref{logprob}), as a function of $a_N$ and $\lambda_N$.
 If $a_N$ is replaced by $a_N'$ given above, one finds
that for any sequence $\lambda_N$
$$A_{Nk}(a_N,\lambda_N)/A_{Nk}(a_N',\lambda_N)\to 1$$
at least as fast as $\eta_N\ln N$ tends to zero.
Therefore equation (\ref{res3.4}) can be replaced by
\be
P_{N,a_N'}(\frac{n_0}{N}\geq \lambda_N)
=e^{-(1+\zeta_N)A_{N1}(a_N,\lambda_N)}
\ee
where $\zeta_N\to 0$. Because the results of Theorem \ref{BECharmfree} grouped under points I and II
were derived from equation (\ref{res3.4}) and did not depend on the particular way $\epsilon_N$ tended to
zero, the modified scaling will provide the same outcome. Although we do not use it, we note
that the free energy also will be the same.

\vspace{2mm}
As a corollary we obtain
\begin{coro}\label{coroGBEC}
Let $S=S(N)=o(N/\ln N)$. For $a<1$
\be\label{a<1}
\lim_{N\to\infty}N^{-a}\langle n_0\rangle_{N+S,a_N}=1.
\ee
\end{coro}

\noindent
{\em Proof.} Because $\lim (N+S)^{a}/N^a=1$,
\[
\lim_{N\to\infty}N^{-a}\langle n_0\rangle_{N+S,a_N}=
\lim_{N\to\infty}N^{-a}\langle n_0\rangle_{N,a_N'}
\]
with
\be
a_N'=\frac{1+(\ln N)^{-1}\ln(1-S/N)}{1-S/N}\ a_N\equiv (1+\eta_N)a_N.
\ee
One can easily verify that $\eta_N\ln N\to 0$. Thus, the result follows by applying
Proposition \ref{propGBEC}.

\begin{lemma}\label{trivilemma}
For any noninteracting Bose gas with a one-particle spectrum $\{\veps_j\}$
the average occupation numbers in the $N$-particle canonical ensemble are strictly decreasing with
the energy,
\be
\langle n_j\rangle_N < \langle n_i\rangle_N\qquad{\rm if}\qquad\veps_i<\veps_j.
\ee
\end{lemma}

\noindent
{\it Proof.} Let $x_i=\exp(-\beta\veps_i)$ and let $Z_N$ and $Z_{N,i,j}$ denote the $N$-particle
partition functions of the full system and of the system with missing levels $i$ and $j$, respectively.
\bea
Z_N[\langle n_i\rangle_N-\langle n_j\rangle_N]=\sum_{l,m}(m-l)x_i^mx_j^lZ_{N-l-m,i,j}
=\sum_{0\leq l<m\leq N}(m-l)[x_i^mx_j^l-x_i^lx_j^m]Z_{N-l-m,i,j}\\
=\sum_{0\leq l<m\leq N}(m-l)(x_ix_j)^l[x_i^{m-l}-x_j^{m-l}]Z_{N-l-m,i,j}>0.
\eea

\begin{theorem}\label{thm4}
For any $a\equiv\hbar\omega\beta>0$ the scaling $a_N=a\ln N/N$ leads to a complete generalized
Bose-Einstein condensation. For $a>a_c=1$ there is no fragmented condensation but the generalized
condensate on the levels $1,2,\ldots$ is at its critical point. In particular:\\
(i) For any $a>0$
\be\label{complete}
\lim_{N\to\infty}\frac{1}{N}\sum_{j<2/a_N}\langle n_j\rangle_{N,a_N}=1.
\ee
(ii) For $a<1$
\be\label{niabove}
\lim_{N\to\infty}N^{-a}\langle n_i\rangle_{N,a_N}=1\quad
\mbox{if}\quad i=o\left(\frac{N^{1-a}}{\ln^2 N}\right).
\ee
(iii) For $a>1$
\be\label{ni1}
\lim_{N\to\infty}N^{-1}\langle n_i\rangle_{N,a_N}=0\quad\mbox{if}\quad i\geq 1
\ee
\phantom{aaaa}but
\be\label{n1eta}
\lim_{N\to\infty}N^{-\eta}\langle n_1\rangle_{N,a_N}=\infty\quad\mbox{for any}\quad\eta<1.
\ee
\end{theorem}

\vspace{2mm}
\noindent
{\it Proof.}\\
(i) Let $J=2/a_N$. We shall prove that
\bea\label{probGBEC}
\ln P_{N,a_N}\left(\sum_{j\geq J}n_j=m\right)\leq -\left(2-\frac{\pi^2}{12}\right)
m\quad{\rm if}\quad m\geq J\ .
\eea
Then for $b=e^{-2+\pi^2/12}$
\bea
\sum_{m\geq J}mP_{N,a_N}\left(\sum_{j\geq J}n_j=m\right)\leq \sum_{J\leq m\leq N} mb^m
<\left[\frac{J}{1-b}+\frac{b}{(1-b)^2}\right]b^J\rightarrow 0
\eea
as $N$ tends to infinity. Thus for $N$ large enough
\bea
0<N-\sum_{j<J}\langle n_j\rangle_{N,a_N}=
\sum_{j\geq J}\langle n_j\rangle_{N,a_N}
&=&
\sum_{m< J}mP_{N,a_N}\left(\sum_{j\geq J}n_j=m\right)\nonumber\\&+&
\sum_{m\geq J}mP_{N,a_N}\left(\sum_{j\geq J}n_j=m\right)
<J+1\ .
\eea
Dividing by $N$ and taking the limit we obtain (\ref{complete}).

To prove (\ref{probGBEC}), consider the probability on the left side. With the notation $x=e^{-a_N}$
\bea
P_{N,a_N}\left(\sum_{j\geq J}n_j=m\right)=
Q_{N,a_N}^{-1}\left(\sum_{\{n_j\}_{j\geq J}:\sum n_j=m}x^{\sum_{j\geq J}jn_j}\right)
\left(\sum_{\{n_j\}_{j=1}^{J-1}:\sum n_j\leq N-m}x^{\sum_{j=1}^{J-1}jn_j}\right).
\eea
The quantity in the first bracket is $x^{mJ}Q_{m,a_N}$, cf. equation (\ref{Qq}).
The quantity in the second bracket can be bounded by dropping the constraint $\sum n_j\leq N-m$,
\be
\left(\sum_{\{n_j\}_{j=1}^{J-1}:\sum n_j\leq N-m}x^{\sum_{j=1}^{J-1}jn_j}\right)
< \prod_{j=1}^{J-1}\frac{1}{1-x^j}=Q_{J-1,a_N},
\ee
see equation (\ref{Q}). Thus,
\bea
P_{N,a_N}\left(\sum_{j\geq J}n_j=m\right)< x^{mJ}Q_{N,a_N}^{-1}Q_{m,a_N}Q_{J-1,a_N}
\leq x^{mJ}Q_{J-1,a_N}
\eea
because $Q_{m,a_N}$ is increasing with $m$.
Taking the logarithm and using $Ja_N=2$,
\be\label{lnprob}
\ln P_{N,a_N}\left(\sum_{j\geq J}n_j=m\right)<
-2m-\sum_{j=1}^{J-1}\ln (1-x^j).
\ee
Now $-\ln(1-x^j)$ is a positive decreasing function of $j$, therefore
\bea\label{-log}
-\sum_{j=1}^{J-1}\ln (1-e^{-ja_N})&\leq&
-\frac{1}{a_N}\int_{0}^{2}\ln (1-e^{-t})\d t\nonumber\\
&<&-\frac{J}{2}\int_{0}^{\infty}\ln (1-e^{-t})\d t=J\pi^2/12\leq m\pi^2/12
\eea
for $m\geq J$. Sustituting this into equation (\ref{lnprob}) we obtain (\ref{probGBEC}).

\vspace{1mm}
\noindent
(ii) To prove (\ref{niabove}) we remark that because of
\[
\langle n_i\rangle_{N,a_N}\leq\langle n_0\rangle_{N,a_N}\asymp N^a,
\]
cf. Lemma \ref{trivilemma}, it suffices to show that $\lim N^{-a}\langle n_i\rangle_{N,a_N}\geq 1$.
Let $K=i\ln N$; then $K=o(N^{1-a}/\ln N)$. From (\ref{ni-n0})
\bea\label{nigeq}
\langle n_i\rangle_{N,a_N}
&\geq&
\sum_{M=0}^{KN^a} \langle n_0\rangle_{N-M,a_N}P_{N,a_N}\left(\sum_{j=0}^{i-1}n_j=M\right)
\geq
\langle n_0\rangle_{N-\lfloor KN^a\rfloor,a_N}
P_{N,a_N}\left(\sum_{j=0}^{i-1}n_j\leq KN^a\right)\nonumber\\
\eea
where we have used the monotonic increase of $\langle n_0\rangle_{k,a_N}$ with $k$, derived in
\cite{S3}. By Corollary \ref{coroGBEC} equation (\ref{nigeq}) implies
\be
\lim_{N\to\infty}N^{-a}\langle n_i\rangle_{N,a_N}\geq 1-\lim_{N\to\infty}
P_{N,a_N}\left(\sum_{j=0}^{i-1}n_j\geq KN^a\right)\ .
\ee
We show that the limit on the right-hand side is zero.
\bea
P_{N,a_N}\left(\sum_{j=0}^{i-1}n_j\geq L\right)
&=&P_{N,a_N}\left(\frac{1}{i}\sum_{j=0}^{i-1}n_j\geq \frac{L}{i}\right)
\leq P_{N,a_N}\left(\max\{n_0,\ldots,n_{i-1}\}\geq \frac{L}{i}\right)\nonumber\\
&\leq&\sum_{j=0}^{i-1}P_{N,a_N}\left(n_j\geq \frac{L}{i}\right)
=\sum_{j=0}^{i-1}Q_{N,a_N}^{-1}\sum_{m\geq L/i}x^{mj}
\sum_{\{n_k\}_{k\neq j}:\sum n_k=N-m}x^{\sum kn_k}
\nonumber\\
&\leq& \sum_{j=0}^{i-1}x^{Lj/i}P_{N,a_N}(n_0\geq L/i,n_j=0)
\leq P_{N,a_N}(n_0\geq L/i)\sum_{j=0}^{i-1}x^{Lj/i}\ .\nonumber\\
\eea
For $L=KN^a=iN^a\ln N$ we can apply (\ref{specasym}) with $\lambda_N=\ln N/N^{1-a}$.
We find
\bea
P_{N,a_N}\left(\sum_{j=0}^{i-1}n_j\geq KN^a\right)
&\leq& \exp\left\{-\frac{1}{N^{a}}\frac{e^{a\ln^2N/N^{1-a}}-1}{e^{a\ln N/N}-1}(1+\epsilon_N)\right\}
\sum_{j=0}^{i-1}x^{Lj/i}\nonumber\\
&<& N^{-(1+\epsilon_N)}i\to 0
\eea
which proves the assertion.

\vspace{1mm}
\noindent
(iii) Choose any $\delta$ such that $1<\delta<a$. Let $K=(\delta/a)N$ and for the sake of notational
simplicity suppose that $K$ is integer. From equation (\ref{ni-n0})
\be
\langle n_1\rangle_{N,a_N}\leq
\langle n_0\rangle_{K,a_N}P_{N,a_N}(n_0\geq N-K)+N P_{N,a_N}(n_0<N-K).
\ee
Now $N=(a/\delta)K$, thus
\[
a_N=\delta\ \frac{\ln K+\ln(a/\delta)}{K}<(2\delta-1)\frac{\ln K}{K}
\]
if $N$ is large enough.
Because $\langle n_0\rangle_{K,a}$ is an increasing function of $a$ (or of $\beta$, see \cite{S3}),
\bea
0\leq
\lim_{N\to\infty}\frac{1}{N}\langle n_1\rangle_{N,a_N}
\leq
\frac{\delta}{a}\left[1-\frac{1}{2\delta-1}\right]
+\lim_{N\to\infty}P_{N,a_N}(n_0<N-K)\phantom{aaaaaaaaaaaaaaaaa}
\nonumber\\
=\frac{\delta}{a}\left[1-\frac{1}{2\delta-1}\right]
+1-\lim_{N\to\infty}P_{N,a_N}\left(\frac{n_0}{N}\geq 1-\frac{\delta}{a}\right)
=\frac{\delta}{a}\left[1-\frac{1}{2\delta-1}\right]
\eea
where we have used equations (\ref{mean}) and (\ref{step}).
Because this holds for any $\delta>1$, letting $\delta$ tend
to 1 we obtain (\ref{ni1}) for $i=1$. For $i>1$ (\ref{ni1}) results by applying Lemma \ref{trivilemma}.

To prove (\ref{n1eta}) choose any $\eta<1$. From equation (\ref{ni-n0})
\bea
\langle n_1\rangle_{N,a_N}\geq \sum_{M>K}\langle n_0\rangle_{M,a_N} P_{N,a_N}(n_0=N-M)
\geq
\langle n_0\rangle_{K,a_N} [1-P_{N,a_N}(n_0\geq N-K)].
\eea
Let $\delta=\frac{1}{2}(\eta+1)$, then $\delta<1$. Define $K=(\delta/a)N$. By (\ref{step}) now
\[
\lim_{N\to\infty}P_{N,a_N}(n_0\geq N-K)=\lim_{N\to\infty}P_{N,a_N}(n_0/N\geq 1-\delta/a)=0.
\]
Because $a_N>\delta\ln K/K$, with equation (\ref{meann0a<1}) we obtain
\bea
\lim_{N\to\infty}\frac{1}{N^\eta}\langle n_1\rangle_{N,a_N}
\geq
\left(\frac{\delta}{a}\right)^\eta\lim_{K\to\infty}K^{1-\delta}\frac{1}{K^\delta}
\langle n_0\rangle_{K,\delta\ln K/K}
=
\left(\frac{\delta}{a}\right)^\eta\lim_{K\to\infty}K^{1-\delta}=\infty
\eea
which finishes the proof of the theorem. We note that
with more effort, using equations (\ref{Gumbel2}) and (\ref{meandetailed})
one could prove the precise asymptotics (\ref{nia>1}) for, at least, any fixed $i\geq 1$.


\subsection{Generalized Bose condensation at all temperatures in the interacting gas}

In this section we investigate the possibility of introducing a nontrivial interaction in the system
without loosing Bose-Einstein condensation. By nontrivial we mean
an interaction providing an energy contribution comparable with or larger than the free
energy of the free system. The Hamiltonian of this latter is
(\ref{harmonic}) with $\gamma_N=\ln N/N$ while that of the interacting gas is
\be\label{Hamint}
H_N=H^0_N+U_N=\sum_{i=1}^N\left[-\h2m\frac{\d^2}{\d x_i^2}+\gamma_NV(\sqrt{\gamma_N}x_i)\right]
+U_N(x_1,\ldots,x_N).
\ee
We return to the notations introduced in Section 2.1.
Mean values with respect to the noninteracting gas in the scaled harmonic trap will thus be denoted by
$\langle\cdot\rangle_{\beta H^0_N}$, in contrast to $\langle\cdot\rangle_{N,a_N}$ used in Sections
3.1 and 3.2. Also, in the case of the spectrally deformed noninteracting gas we apply the notation
$\langle\cdot\rangle_{\beta H^0_N(J,\delta)}$.

\begin{theorem}\label{GBECharmint}
Interacting bosons in a scaled harmonic trap with
\[
\gamma_N=\frac{\ln N}{N}\quad {\rm and}\quad U_N\geq -BN,\ \langle U_N\rangle_{\beta H^0_N}=o(N\ln N)
\]
undergo a complete generalized Bose-Einstein condensation
at all temperatures: For any $\beta>0$ 
\be\label{GBEC}
\lim_{s\to 0}
\lim_{N\to\infty}\frac{1}{N}\sum_{j=0}^{sN}\langle n_j\rangle_{\beta H_N}
=1\ .
\ee
For stable pair interactions,
\[
U_N(x_1,\ldots,x_N)=\sum_{i<j}u_N(x_i-x_j),
\]
$\langle U_N\rangle_{\beta H^0_N}=o(N\ln N)$ and thus (\ref{GBEC}) holds true if $\|u_N\|_1=o(\gamma_N)$.
If $u_N(x)=b_N u(\alpha_Nx)$, the condition reads $\|u\|_1<\infty$ and
$b_N=o(\alpha_N \ln N/N)$.
\end{theorem}

Observe that
the scaling condition $\|u_N\|_1=o(\ln N/N)$ is somewhat weaker than (\ref{uN})
and, again, examples from mean-field type to sharply concentrated interactions
can be obtained.

We note that the discussion of Section 2.2
applies to the case of a fixed harmonic potential. In analogy with Theorem \ref{superharm}
we find that for a {\em fixed} harmonic trap and
scaled interactions satisfying $\langle U_N\rangle_{\beta H^0_N}=o(N^2)$ there is a complete GBEC at all
$\beta>0$. In Theorem \ref{GBECharmint} we conclude that for somewhat weaker interactions the same
holds true even if the frequency of the confining harmonic potential is scaled with $\gamma_N=\ln N/N$.
One could also show without much further ado that for more general scalings $\omega_N=\gamma_N\omega$,
where $\gamma_N\geq\ln N/N$, and $\langle U_N\rangle_{\beta H^0_N}=o(N^2\gamma_N)$ there is a complete
GBEC at all temperatures.

\vspace{2mm}
\noindent
{\it Proof.}\\
We use Lemma \ref{ineq} and Proposition \ref{propgen}. If $J=\lfloor sN\rfloor$ then
$\delta=\veps_J=\hbar\omega\gamma_N(\lfloor sN\rfloor+\frac{1}{2})\approx s\hbar\omega\ln N$ implying
\[\lim\frac{1}{N\delta}[\langle U_N\rangle_{\beta H^0_N}-\inf U_N]=0\]
and therefore
\[
\lim\frac{1}{N}\sum_{j=0}^J\langle n_j\rangle_{\beta H_N}\geq
\lim\frac{1}{N}\sum_{j=0}^J\langle n_j\rangle_{\beta H^0_N(J,\delta)}\ .
\]
 We can conclude by showing that
\be\label{corrineq}
\langle n_j\rangle_{\beta H^0_N(J,\delta)}\geq \langle n_j\rangle_{\frac{1}{2}\beta H^0_N}
\ee
because from Theorem \ref{thm4} it follows that
\[
\lim_{N\to\infty} \frac{1}{N}\sum_{j=0}^J\langle n_j\rangle_{\frac{1}{2}\beta H^0_N}=1
\]
for any $\beta$. To prove (\ref{corrineq}) notice that the spectrum of
$H^0(J,\delta)-\veps_0-\veps_J$ is a part of the spectrum of $\frac{1}{2}(H^0-\veps_0)$. Indeed,
\be
{\rm spec\,}\left\{\frac{1}{2}(H^0-\veps_0)\right\}
=\frac{1}{2}\gamma_N\hbar\omega\times\{0,1,2,\ldots\}
\ee
while
\be
{\rm spec\,}\left\{H^0(J,\delta)-\veps_0-\veps_J\right\}=\frac{1}{2}\gamma_N\hbar\omega\times
\{0,1,2,\ldots,2J,2J+1,2J+3,2J+5,\ldots\}.
\ee
In \cite{S3} we proved the following result.

\begin{lemma}\label{L3.2}
Let $H^0$ and $H^1$ be two one-particle Hamiltonians, both $e^{-\beta H^0}$
and $e^{-\beta H^1}$ trace class and
\[
{\rm spec\,}H^0\subset {\rm spec\,}H^1
\]
where repeated eigenvalues are considered separately. If $\veps_i$ is a common eigenvalue then the
averages of the occupation number $n_i$ in the two noninteracting ensembles satisfy the
inequality
\be\label{3rd}
\langle n_i\rangle_{\beta H^1_N}<\langle n_i\rangle_{\beta H^0_N}\ .
\ee
\end{lemma}
On the left- and right-hand sides of (\ref{3rd})
$n_i$ denotes $n[\varphi^1_i]$ and $n[\varphi^0_i]$,
respectively, where $H^0\varphi^0_i=\veps_i\varphi^0_i$, $H^1\varphi^1_i=\veps_i\varphi^1_i$, and
$\varphi^0_i$ and $\varphi^1_i$ may be different.
If $\eta_j$ are the eigenvalues of $H^1$ not contained in the spectrum of $H^0$ and $x_j=e^{-\beta\eta_j}$
then
\be\label{toL3.4}
\langle n_i\rangle_{\beta H^1_N}=\frac{1}{Z[\beta H^1_N]}\sum_{m=0}^N\langle n_i\rangle_{\beta H^0_{N-m}}
Z[\beta H^0_{N-m}]\sum_{\{k_j\}:\sum k_j=m}\prod_jx_j^{k_j}
\equiv\sum_{m=0}^N\langle n_i\rangle_{\beta H^0_{N-m}}P_m
\ee
where $Z$ denotes partition functions. Because $\sum_{m=0}^NP_m=1$, (\ref{3rd}) follows from
the strict monotonic increase of $\langle n_i\rangle_{\beta H^0_{N}}$ with $N$, proven in \cite{S3}.
We obtain (\ref{corrineq}) and thereby the proof of a complete GBEC 
by replacing $H^0$ with $H^0(J,\delta)-\veps_0-\veps_J$ and
$H^1$ with $\frac{1}{2}(H^0-\veps_0)$ in Lemma \ref{L3.2}. (To avoid confusion we precise that on the
right-hand side of equation (\ref{toL3.4}) $H^0_{N-m}$ is defined with the same one-particle spectrum
for each $m$.)

In the case of pair interactions
\be
\langle u_N\rangle_{\beta H^0_N}=\frac{2}{N(N-1)}\langle U_N\rangle_{\beta H^0_N}
\ee
and to satisfy the condition of the theorem we need $\langle u_N\rangle_{\beta H^0_N}=o(\ln N/N)$.
We show below that
\be\label{uNineq}
|\langle u_N\rangle_{\beta H^0_N}|\leq \frac{\sqrt{2}}{\lambda_\beta}(1+O(\sqrt{\gamma_N})) \|u_N\|_1
\ee
and therefore $\|u_N\|_1=o(\gamma_N)$ is a sufficient condition for GBEC to hold for any $\beta>0$.
Now
\bea\label{uNformula}
\langle u_N\rangle_{\beta H^0_N}&=&Z[\beta H^0_{N}]^{-1}\sum_{g\in S_N}
\int\d x_1\d x_2 u_N(x_1-x_2)\int\d x_3\ldots\d x_N\prod_{j=1}^N\langle x_j|e^{-\beta H^0}|x_{g(j)}\rangle
\nonumber\\
&=&\sum_{g\in S_N}\frac{\Tr U(g)e^{-\beta H^0_N}}
{\sum_{h\in S_N}\Tr U(h)e^{-\beta H^0_N}}
\int\d x_1\d x_2 u_N(x_1-x_2)\mu_{g,\beta}(x_1,x_2)
\eea
where $S_N$ is the group of the permutations of $1,2,\ldots,N$,
$U(g)$ is the unitary representation of the permutation
$g$ in the full $N$-particle Hilbert space and $\Tr$ is the trace in this space.
In the average above there are two kinds of probability measures $\mu_{g,\beta}$.
If 1 and 2 are in different cycles of $g$ then
\be
\mu_{g,\beta}(x,y)=\mu_{\ell_1(g)\beta}(x)\mu_{\ell_2(g)\beta}(y)\equiv
\mu_{\beta_1}(x)\mu_{\beta_2}(y)
\ee
where $\ell_i(g)$ is the length of the cycle of $g$ containing $i$ and
\be
\mu_\beta(x)=\frac{\langle x|e^{-\beta H^0}|x\rangle}{\tr e^{-\beta H^0}}\ .
\ee
If 1 and 2 are in the same cycle of length $\ell$ and $g^j(1)=2$ 
then
\be
\mu_{g,\beta}(x,y)=\frac{\langle x|e^{-j\beta H^0}|y\rangle
\langle y|e^{-(\ell-j)\beta H^0}|x\rangle}{\tr e^{-\ell\beta H^0}}\ .
\ee
In the first case by Fourier transforming and applying Schwarz inequality and Parseval formula we get
\be\label{factorized}
\left|\int\d x_1\d x_2 u_N(x_1-x_2)\mu_{g,\beta}(x_1,x_2)\right|
\leq \|u_N\|_1 \|\mu_{\beta_1}\|\  \|\mu_{\beta_2}\|
\ee
with $\|\mu_\beta\|$ denoting the usual $L^2$ norm of $\mu_\beta$.
In the second case
\bea\label{unfact}
\left|\int\d x_1\d x_2 u_N(x_1-x_2)\mu_{g,\beta}(x_1,x_2)\right|
&\leq&\int\d z |u_N(z)|\int\d y\,\mu_{g,\beta}(z+y,y)\nonumber\\
&\leq& \|u_N\|_1\sup_z\int\d y\,\mu_{g,\beta}(z+y,y).
\eea
At this point we recall that (\ref{int_to_pot}) involves a transformation between two Hilbert spaces,
that of $L^2$ functions of the variables $\x_i$ and $\y_i=\alpha_N \x_i$, respectively.
Previously we
have only been interested in properties depending on the spectrum of $\beta H^0$.
Because $\beta$ and $\omega$ appeared in a single dimensionless combination $a=\hbar\omega\beta$,
$\gamma_N$ could be
considered to multiply $a$ and yield $a_N=a\gamma_N$. Here we are estimating functions and have to
decide which space to work in. In accordance with equations (\ref{harmonic}) and (\ref{Hamint})
we choose to scale the potential which amounts to replace $\omega$ by
$\omega_N=\omega\gamma_N$ and $a$ by $a_N=\hbar\omega_N\beta$ and to keep $\beta$ as an independent
unscaled variable.

We use Mehler formula \cite{R}
\be
\langle x|e^{-\beta H^0}|y\rangle=
\left[\frac{m\omega_N}{2\pi\hbar\sinh a_N}\right]^{\frac{1}{2}}
\exp\left\{-\frac{m\omega_N(x^2+y^2)}{2\hbar\tanh a_N}+\frac{m\omega_N xy}{\hbar\sinh a_N}\right\}
\ee
to obtain
\be
\mu_\beta(x)=\sqrt{s/\pi}\exp(-sx^2)\qquad
s=s(\beta)=\frac{m\omega_N}{\hbar}\tanh \hbar\omega_N\beta/2\ .
\ee
Because
\be
\frac{\partial\mu_\beta(x)}{\partial s}=\mu_\beta(x)[(2s)^{-1}-x^2]
=\mu_\beta(x)[\langle x^2\rangle_{\mu_\beta}-x^2]\ ,
\ee
$\|\mu_\beta\|$ is an increasing function of $s$ and, therefore, of $\beta$:
\bea
\frac{\partial}{\partial s}\int\mu_\beta^2 \d x&=&2\int_{x^2<1/2s}\mu_\beta(x)
\frac{\partial\mu_\beta}{\partial s}(x)\d x+
2\int_{x^2>1/2s}\mu_\beta(x)
\frac{\partial\mu_\beta}{\partial s}(x)\d x\nonumber\\ &>&
2\mu_\beta(\sqrt{1/2s})\left[
\int_{x^2<1/2s}
\frac{\partial\mu_\beta}{\partial s}(x)\d x+
\int_{x^2>1/2s}
\frac{\partial\mu_\beta}{\partial s}(x)\d x\right]=0\nonumber\ .
\eea
Here we used that $\mu_\beta(x)$ is a decreasing function of $|x|$ and $\int\mu_\beta \d x=1$.
Thus, we can bound (\ref{factorized}) by inserting the largest possible $\beta_1$ and $\beta_2$ on the
right-hand side. Because
\[
\max_g \ell_i(g)\beta=(N-1)\beta<N\beta\ ,
\]
we find
\bea\label{type1}
\|\mu_{\beta_1}\|\  \|\mu_{\beta_2}\|
&<&\|\mu_{N\beta}\|^2
=\frac{s(N\beta)}{\pi}\int e^{-2s(N\beta)x^2}\d x=\sqrt{s(N\beta)/2\pi}
=\left[\frac{m\omega_N}{2\pi\hbar}\tanh\hbar\omega_N N\beta/2\right]^{1/2}\nonumber\\
&<&\left[\frac{m\omega_N}{2\pi\hbar}\right]^{1/2}=O(\sqrt{\gamma_N})\ .
\eea
To estimate the right-hand side of (\ref{unfact}), observe that
\[
\nu(z)=\int\mu_{g,\beta}(z+y,y)\d y
\]
is normalized. One then obtains
\be
\nu(z)=\sqrt{D/\pi}e^{-Dz^2} \qquad D=\frac{m\omega_N}{4\hbar}
\left(\frac{1}{\tanh ja_N}+\frac{1}{\tanh (\ell-j)a_N}+\frac{1}{\sinh ja_N}+\frac{1}{\sinh (\ell-j)a_N}\right)
\ee
and thus $\sup_z\nu(z)=\nu(0)=\sqrt{D/\pi}$. Now $D$ attains its maximum for $\ell=2$,
\[
\max_{\ell\geq 2} D=\frac{m\omega_N}{2\hbar}\left(\frac{1}{\tanh a_N}+\frac{1}{\sinh a_N}\right)
=\frac{m}{\hbar^2\beta}+O(\gamma_N^2)
=\frac{2\pi}{\lambda_\beta^2}+O(\gamma_N^2)\ .
\]
Thus for the permutations containing 1 and 2 in the same cycle we have the uniform upper bound
\be\label{type2}
\sup_z\int\mu_{g,\beta}(z+y,y)\d y\leq \frac{\sqrt{2}}{\lambda_\beta}+O(\gamma_N^2)\ .
\ee
Finally, we obtain the inequality (\ref{uNineq}) by substituting
(\ref{type1}) and (\ref{type2}) into (\ref{factorized}) and (\ref{unfact}), respectively,
and using these latter to bound the integrals in (\ref{uNformula}).
The consequence $b_N=o(\alpha_N\gamma_N)$
for scaled pair interactions is trivial.
By this we finished the proof of the theorem.

\section{Summary}

This paper is an extension of our earlier study \cite{S} of Bose-Einstein condensation of trapped
Bose gases. We have concentrated on two problems. The first, treated in Section 2,
concerns bosons in a fixed trap
interacting with unscaled interactions; that is, neither the confining potential nor the pair interaction
between two particles depends on the number of particles.
Harmonic or weaker trap potentials combined with unscaled
interactions are too difficult to deal with. However,
if the confining potential increases faster than quadratically at infinity, in one dimension we can prove
a complete generalized Bose-Einstein condensation at all temperatures. This means that as $N$ tends to
infinity all but a vanishing fraction of particles will be distributed over a set of one-particle
states whose number is asymptotically negligible compared with $N$.
The result does not provide a
more precise information about the number of one-particle states carrying the condensate.
The generalized condensation may eventually prove to be normal and non-fragmented.

The second problem we have investigated in this paper is the condensation of bosons in a one-dimensional
scaled harmonic trap. In a previous work \cite{KvD} it was shown that for a particular scaling, when
the oscillator frequency $\omega$ is replaced by $\omega\ln N/N$, Bose-Einstein condensation
occurs with a phase transition in the noninteracting gas. Sections 3.1 and 3.2 have been devoted to a
detailed study of this system. Our most interesting finding is that the occupation of the excited states
is anomalously large and results in a complete generalized Bose-Einstein condensation
at all temperatures, superimposed on the normal Bose-condensation below the critical temperature.
In Section 3.3 we have studied the same system in the case when there is also a suitably scaled interaction
among the particles. We have shown that the complete GBEC is preserved by the interaction without
being able to prove that the phase transition also persists.

\vspace{3mm}
\noindent
{\bf\large Acknowledgment}\\
This work was partially supported by the Hungarian Scientific Research Fund (OTKA) Grants T42914
and T46129 and the Center of Excellence Grant ICA 1-CT-2000-70029.


\begin{thebibliography}{99}
\bibitem{Gir}
M. Girardeau: {\em Relationship between systems of impenetrable bosons and fermions in one dimension.}
J. Math. Phys. {\bf 1} 516-523 (1960)
\bibitem{Sch}
T. D. Schultz: {\em Note on the one-dimensional gas of impenetrable point-particle bosons.}
J. Math. Phys. {\bf 4} 666-671 (1963)
\bibitem{Len}
A. Lenard: {\it Momentum distribution in the ground state of the one-dimensional
system of impenetrable bosons.} J. Math. Phys. {\bf 5} 624-637 (1964)
\bibitem{H}
P. C. Hohenberg: {\em Existence of long-range order in one and two dimensions.}
Phys. Rev. {\bf 158} 383-386 (1967)
\bibitem{BM}
M. Bouziane and Ph. Martin: {\em Bogoliubov inequality for unbounded operators and the
Bose gas.} J. Math. Phys. {\bf 17} 1848-1851 (1976)
\bibitem{G2}
M. D. Girardeau: {\em Broken symmetry and generalized Bose condensation in restricted geometries.}
J. Math. Phys. {\bf 10} 993-998 (1969)
\bibitem{B}
N. N. Bogoliubov: {\em Quasi-averages in problems of statistical mechanics.} Dubna report D-781 (1961),
in Russian
\bibitem{LL}
E. H. Lieb and W. Liniger: {\em Exact analysis of an interacting Bose gas. I. The general solution of the
ground state.} Phys. Rev. {\bf 130} 1605-1616 (1963)
\bibitem{Hal}
F. D. M. Haldane: {\em Effective harmonic-fluid approach to low-energy properties of one-dimensional
quantum fluids.} Phys. Rev. Lett. {\bf 47} 1840-1843 (1981)
\bibitem{CTW}
D. B. Creamer, H. B. Thacker and D. Wilkinson:
{\em A study of correlation functions for the delta-function Bose gas.} Physica {\bf 20D} 155-186 (1986)
\bibitem{KBI}
V. E. Korepin, N. M. Bogoliubov and A. G. Izergin:
{\em Quantum inverse scattering method and correlation functions.} Cambridge University Press (1993)
Ch. XVIII.2
\bibitem{Col}
S. Coleman: {\em There are no Goldstone bosons in two dimensions.}
Commun. Math. Phys. {\bf 31} 259-264 (1973)
\bibitem{PS}
L. Pitaevskii and S. Stringari: {\em Uncertainty principle, quantum fluctuations, and
broken symmetries.} J. Low Temp. Phys. {\bf 85} 377-388 (1991)
\bibitem{S}
A. S\"ut\H o: {\em Thermodynamic limit and proof of condensation for trapped
bosons.} J. Stat. Phys. {\bf 112} 375-396 (2003)
\bibitem{Sym}
K. Symanzik: {\em Proof and refinements of an inequality of Feynman}.
J. Math. Phys. {\bf 6} 1155-1156 (1965),
\bibitem{LS}
E. H. Lieb and R. Seiringer: {\em Proof of Bose-Einstein condensation for dilute trapped gases.}
Phys. Rev. Lett. {\bf 88} 170409 (2002)
\bibitem{Sei}
R. Seiringer: {\em Ground state asymptotics of a dilute, rotating gas.}
J. Phys. A: Math. Gen. {\bf 36} 9755-9778 (2003)
\bibitem{DGPS}
F. Dalfovo, S. Giorgini, L. P. Pitaevskii and S. Stringari:
{\em Theory of Bose-Einstein condensation in trapped gases.}
 Rev. Mod. Phys. {\bf 71} 463-512 (1999)
\bibitem{LSY1}
E. H. Lieb, R. Seiringer and J. Yngvason: {\em Bosons in a trap: A rigorous derivation of the
Gross-Pitaevskii energy functional.}
Phys. Rev. A {\bf 61} 043602 (2000)
\bibitem{Goe}
A. G\"orlitz, J. M. Vogels, A. E. Leanhardt, C. Raman, T. L. Gustavson, J. R. Abo-Shaeer,
A. P. Chikkatur, S. Gupta, S. Inouye, T. Rosenband and W. Ketterle:
{\em Realization of Bose-Einstein condensates in lower dimensions.}
Phys. Rev. Lett. {\bf 87} 130402 (2001)
\bibitem{Bou}
P. Bouyer, J. H. Thywissen, F. Gerbier, M. Hugbart, S. Richard, J. Retter and A. Aspect:
{\em One-dimensional behaviour of elongated Bose-Einstein condensates.}
Quantum gases in low dimensions. Les Houches Proceedings (2003)
\bibitem{LSY}
E. H. Lieb, R. Seiringer and J. Yngvason: {\em One-dimensional bosons in three-dimensional traps.}
Phys. Rev. Lett. {\bf 91} 150401 (2003)
\bibitem{PSW}
D. S. Petrov, G. V. Shlyapnikov and J. T. M. Walraven:
{\em Regimes of quantum degeneracy in trapped 1D gases.}
Phys. Rev. Lett. {\bf 85} 3745-3749 (2000)
\bibitem{Ger}
F. Gerbier, J. H. Thywissen, S. Richard, M. Hugbart, P. Bouyer and A. Aspect:
{\em Momentum distribution and correlation function of quasicondensates in elongated traps.}
Phys. Rev. A {\bf 67} 051602(R) (2003)
\bibitem{KvD}
W. Ketterle and N. J. van Druten: {\em Bose-Einstein condensation of a finite
number of particles trapped in one or three dimensions.} Phys. Rev. A {\bf 54}
656-660 (1996)
\bibitem{ZB}
V. A. Zagrebnov and J.-B. Bru: {\em The Bogoliubov model of weakly imperfect Bose gases.}
Phys. Rep. {\bf 350} 291-434 (2001)
\bibitem{Tit}
E. C. Titchmarsch: {\em Eigenfunction expansions I.} Oxford at the Clarendon Press (1962) Ch. VII
\bibitem{BP}
E. Buffet and J. Pul\'e: {\em Fluctuation properties of the imperfect Bose gas.}
J. Math. Phys. {\bf 24} 1608-1616 (1983)
\bibitem{S3}
A. S\"ut\H o: {\em Correlation inequalities for noninteracting Bose gases.} 
J. Phys. A: Math. Gen. {\bf 37} 615-621 (2004)
\bibitem{R}
G. Roepstorff: {\em Path Integral Approach to Quantum Physics} (Springer, Berlin, 1994)
\end{thebibliography}
\end{document}